\documentclass[aps,pra,reprint, superscriptaddress,showpacs]{revtex4-2}

\usepackage{dsfont}
\usepackage{dcolumn}
\usepackage{amsmath}
\usepackage{amssymb}
\usepackage{physics}
\usepackage{quantikz}
\usepackage{float}
\usepackage{graphicx}
\usepackage{bm}
\usepackage[colorlinks = true, linkcolor = blue, urlcolor = blue, citecolor = red, anchorcolor = blue]{hyperref}
\usepackage{verbatim}
\usepackage[dvipsnames]{xcolor} 
\usepackage{soul} 
\usepackage{enumitem}

\usepackage[makeroom]{cancel} 

\newcommand{\red}[1]{{\color{BrickRed}{#1}}}

\begin{document}
\title{Optimizing resource allocation for accuracy in noisy variational quantum algorithms}
\author{Harshit Verma}
 \email{harshit.verma@cnrs.fr}
  \affiliation{MajuLab, CNRS-UCA-SU-NUS-NTU International Joint Research Laboratory, 117543 Singapore,}
 \affiliation{Centre for Quantum Technologies, National University of Singapore, 117543 Singapore.}

\author{Thomas Ayral}
\affiliation{CPHT, CNRS, École Polytechnique, IP Paris, F-91128 Palaiseau, France,}
\affiliation{Eviden Quantum Laboratory, 78340 Les Clayes-sous-Bois, France.}

\author{Alexia Auffèves}
  \affiliation{MajuLab, CNRS-UCA-SU-NUS-NTU International Joint Research Laboratory, 117543 Singapore,}

\affiliation{Centre for Quantum Technologies, National University of Singapore, 117543 Singapore.}

\author{Robert Whitney}
\affiliation{Université Grenoble Alpes, CNRS, LPMMC, 38000 Grenoble, France.}

\begin{abstract}
For quantum algorithms to achieve their full potential, we need methodologies to optimize them, such as reaching a given output accuracy with minimal resource costs. Here, we develop such a methodology for a class of Noisy Intermediate-Scale Quantum (NISQ) algorithms.
We leverage simulations of a Variational Quantum Eigensolver (VQE) to propose a phenomenological 
model of such algorithms that captures the complex relationship between algorithmic accuracy, 
algorithmic resource costs, and the noise that exists in realistic quantum hardware. For this, we take the algorithmic resource cost to be the total number of quantum gate-operations in the algorithm; minimizing this cost typically makes the algorithm faster and more energy-efficient. We consider the subtle trade-off between quantum circuit size (small circuits are too imprecise, but large ones are too noisy), and the number of iterations of that quantum circuit for the full algorithm to sufficiently converge. Using a noise-metric-resource methodology, we identify the sweet spot (of circuit size versus iterations) that minimizes the algorithmic resource costs for a desired algorithm accuracy. 
It also gives the circuit size that maximizes algorithm accuracy for a fixed resource cost.  Our methodology provides a practical guideline for near-term deployment of variational algorithms on realistic noisy hardware, including hardware that uses error mitigation.
\end{abstract}
\maketitle

\section{Introduction}

\begin{figure}[t]
    \centering
    \includegraphics[width = \columnwidth]{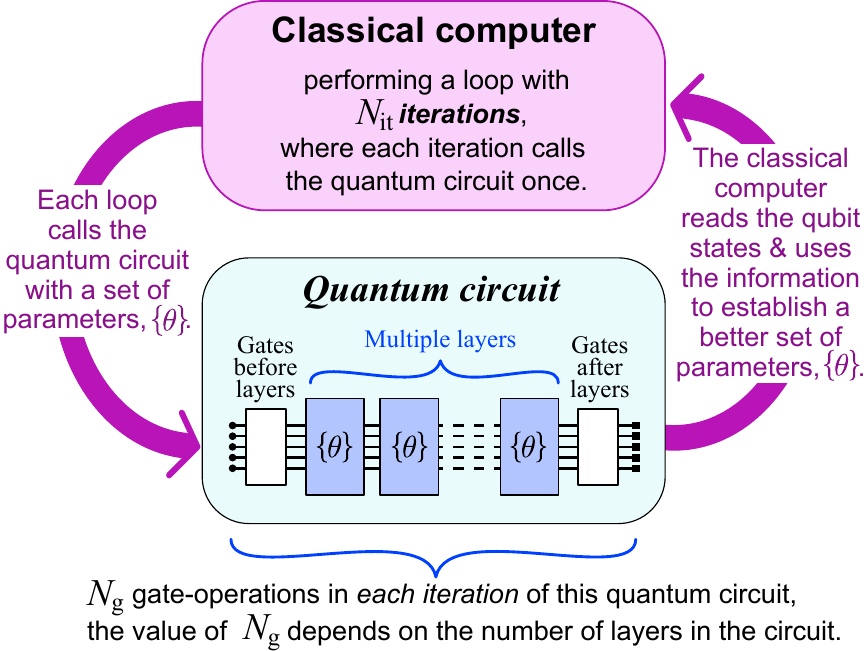}
    \caption{A typical Noisy Intermediate-Scale Quantum (NISQ) algorithm, in this case a Variational Quantum Eigensolver (VQE) for approximating the ground-state energy ($E_\textrm{gs}$) of a given Hamiltonian. A classical optimizer begins with a randomly chosen initial set of variational parameters $\{\theta\}$. A parameterized quantum circuit (ansatz) is then prepared and executed with these parameters, and the expectation value of the Hamiltonian's energy is estimated from measurements of the resulting state. The classical optimizer uses this energy estimate to update $\{\theta\}$ according to a chosen optimization strategy. This loop repeats for $N_{\rm it}$ iterations, ideally converging to a set of parameters $\{\theta^*\}$ that yields a close approximation to the true ground-state energy.
          }
    \label{fig:basic}
\end{figure}

Variational Quantum Eigensolvers (VQE) \cite{Tilly2022} are leading algorithms for Noisy Intermediate-Scale Quantum (NISQ) computing \cite{Preskill2018, Bharti2022}. They are iterative algorithms in which each iteration requires a quantum computer to run a quantum circuit containing a certain set of gate operations, see Fig.~\ref{fig:basic}.
In principle, such an algorithm's accuracy is improved by increasing the number of gate-operations in the quantum circuit. However, all quantum circuits implemented in the near-term will exhibit errors induced by noise; while error mitigation techniques can alleviate these errors \cite{Kurita2023synergeticquantum}, they do not eliminate them. 
Such errors accumulate as the number of operations in the quantum circuit increases \cite{Fontana2020, de2023limitations, Wang2021}, and this determines the largest circuit that can produce useful results. At the same time, the physical resource cost of an algorithm, specifically the time taken and energy consumed, grow with both the number of gate-operations in the quantum circuit and the number of iterations of that circuit that are required to achieve a certain algorithmic accuracy. Key questions then arise. What is the circuit size  with maximizes the accuracy in the presence of noise?
What is required to minimize the resources of time and energy that are necessary for a desired algorithm accuracy?

This mandates the development of methods to minimize the resources consumed by quantum algorithms for a desired algorithm accuracy. Such minimizations are required for all aspects of the quantum computers, both hardware and software \cite{energy_advantage}.
Without such methods, near-term quantum computer are likely to be slower and consume more energy than classical computer when solving problems of practical interest.

In this work, we propose a methodology for minimizing the total algorithmic resources required to implement a VQE algorithm in a manner that guarantees a desired accuracy of the algorithm's output.  For given hardware and given (desired) algorithm accuracy, this will typically minimize both the time taken by the algorithm and the energy consumed by the algorithm.
This methodology goes beyond circuit-level optimization to holistically minimize the algorithmic resource cost of VQE algorithm, by accounting for the interplay between performance, noise and resource costs, within a Metric-Noise-Resource framework \cite{marco2023}. For this, we propose and analyze a phenomenological model that quantifies the relations between a metric (the algorithm's accuracy), the noise (whose effect is reduced but not eliminated with error mitigation techniques), and the algorithmic resources (number of gates in the quantum circuit and number of iterations of that circuit).  Our approach parallels recent work in classical machine learning, where predictive scaling relations relate model performance to computational resources \cite{kaplan2020scaling, hoffmann2022training}. However, the specific nature of noise-induced errors in quantum algorithms is a crucial aspect of our approach.

To be more precise about our optimization strategy, one must recall that a typical VQE algorithm works as sketched in Fig~\ref{fig:basic}. Its goal is to find a good approximation of the ground-state energy of a quantum system of interest. It works iteratively, by repeatedly refining the parameters of a quantum circuit until that circuit converges to a quantum state whose energy closely approximates that ground-state energy. This process involves two algorithmic parameters: the size of the circuit, set by the number of gates it contains $N_\textrm{g}$, and the number of iterations of that circuit $N_\textrm{it}$.  Typically, for any given number of qubits and given hardware, both the time taken by the algorithm and the energy consumed by the algorithm are dependent on $\Delta$, where $\Delta = N_\textrm{g}N_\textrm{it}$. Hence, we take $\Delta$ to be the algorithmic resource that we want to minimize.
The main algorithmic control parameters that we can adjust to optimize the algorithm are $N_\textrm{g}$ and $N_\textrm{it}$.
In the absence of noise, increasing $N_\textrm{g}$ adds more parameters to the quantum circuit, allowing it to prepare states that span a larger portion of the Hilbert space, so iterations of this circuit will converge to a higher accuracy result.
In the presence of noise, this is only true when  $N_\textrm{g}$ is small enough; if the circuit is too large, there is a significant probability of errors in each iteration, making convergence to an accurate result less likely.

Previous works have explored minimizing the resource usage of Variational Quantum Eigensolvers, including  grouping measurements \cite{clique_measure, Yen2023, Nakaji2023measurement}, parallelization \cite{Cattelan2025, parallel1}, new optimization techniques \cite{PhysRevResearch.5.043136, Ostaszewski_2021}, and efficient hardware encoding \cite{Bravyi_2002}. Alternative directions such as adaptive ansatz  \cite{Grimsley2019, Ryabinkin2018}, over-parameterization \cite{You2022, Haug2021}, and quantum architecture search \cite{Du_2022} also aim to reduce resource cost through improved circuit design. For other examples in other NISQ computing and longer-term large-scale quantum computing, see \cite{gidney2021factor,astra, qpe_tradeoff, fellousasiani2024magicstatesrarelyimportant,gidney2025factor2048bitrsa}. 
However, in many cases, the circuit depth and number of iterations are treated as fixed parameters chosen heuristically (with the risk of suboptimal choices) or through grid searches
(with the significant overhead of running the circuit a vast number of times just to find a good choice) \cite{thesis_hyper, hyper_params, molecular_hyper_params, Bonet_Monroig_2023, Herbst_2024, PhysRevResearch.6.023009, Mihalikova2025}.  
Here, instead we develop phenomenological scaling relations. We initially explain these relations for the simplest noise model 
(global depolarizing noise) without error mitigation, but we then show that 
they also apply to a realistic noise model (taken from experimental hardware) with 
error mitigation.  These phenomenological scaling relations 
can either provide a better intuition for heuristic choices of circuit parameters, or 
be used (as we show here) to establish optimal parameters from data given by a modest number of runs of the circuit.



\begin{figure}[t]
    \centering
    \includegraphics[width = \columnwidth]{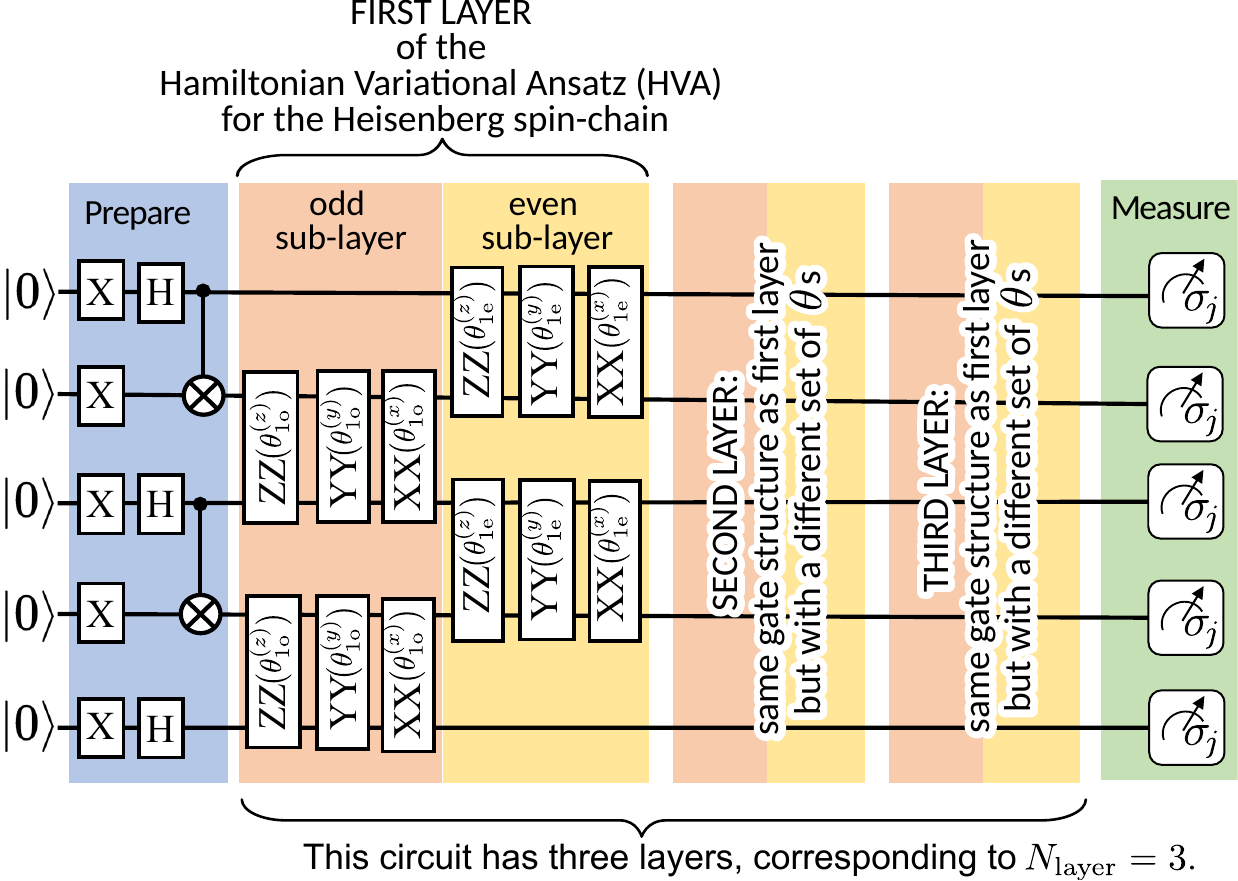}
          \caption{Abstract representation of the quantum  circuit implementing a VQE algorithm for the Heisenberg model in Eq.~\eqref{eq:heisen}, using a slight variant of the Hamiltonian Variational Ansatz (HVA) in Fig.~2 of Ref.~\cite{Wiersema2020}. The circuit shown here has three layers, $N_\textrm{layers}=3$, but we will consider circuits with arbitrary $N_\textrm{layers}$.  The $q$th layer has the following 6 classical variational parameters: $\theta_{q\textrm{o}}^{(z)}$, $\theta_{q\textrm{e}}^{(z)},\theta_{q\textrm{o}}^{(y)}$, $\theta_{q\textrm{e}}^{(y)}$, $\theta_{q\textrm{o}}^{(x)}$ and  $\theta_{q\textrm{e}}^{(x)}$.
          The two qubit gates here are $\textrm{XX}(\theta)= e^{-\red{i}\sigma_x \otimes \sigma_x \theta}$, $\textrm{YY}(\theta)= e^{-\red{i}\sigma_y \otimes \sigma_y \theta}$ and  $\textrm{ZZ}(\theta)= e^{-\red{i}\sigma_z \otimes \sigma_z \theta}$. Some of the variational angles are shared between multiple gates as per the ansatz prescription. Information about the state is obtained through the qubit measurements, and this data is fed to the classical optimizer as shown in Fig.~\ref{fig:basic}. In this work a transpiled version of this circuit is used (see Fig.~\ref{fig:HVA}).
          }
    \label{fig:HVA_abstract}
\end{figure}

\begin{figure*}
    \centering
    \includegraphics[width = \linewidth]{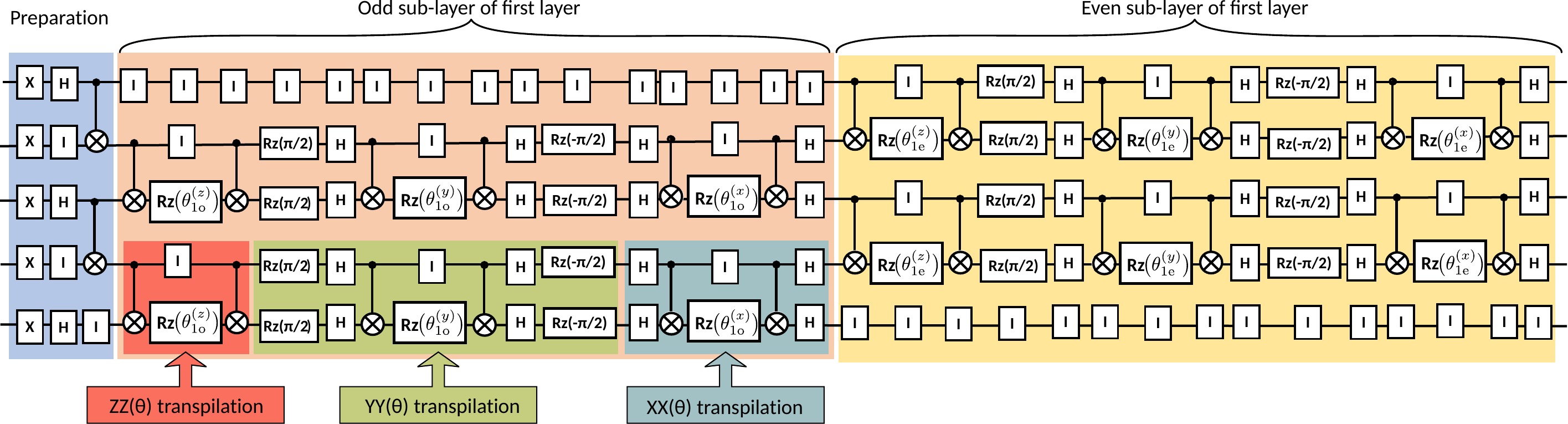}
           \caption{ The transpiled quantum circuit that we considered (showing only the first layer of that circuit). This transpiled circuit is given by taking the quantum circuit in  Fig.~\ref{fig:HVA_abstract}, while assuming that $\textrm{ZZ}(\theta)$,  $\textrm{YY}(\theta)$ and  $\textrm{XX}(\theta)$) are not native gates of the hardware, and so they must be have transpiled into native gates. Here we assume those native gates are $Ry$, $Rz(\theta)$ and CNOT gates. Individual colored boxes show how $\textrm{ZZ}(\theta)$,  $\textrm{YY}(\theta)$ and  $\textrm{XX}(\theta)$ are transpiled into such native gates. Each identity gates is indicated by ``\,\textsf{I}\,''. We observe that there are 150 gates per layer (including 42 identity gates), when we count gates in the manner explained above  Eq.~\eqref{eq:N_g}.
           }
    \label{fig:HVA}
\end{figure*}

When judging if a machine performs a task efficiently, one must choose a definition of efficiency (for example, a supercomputer's energy efficiency is given as its GFLOPS per Watt \cite{Green500}). 
We use the Metric–Noise–Resource (MNR) framework \cite{marco2023} as a unifying logic for maximizing the efficiency of a noisy quantum computer. 
In that framework, one identifies a suitable metric of performance of the task of interest, $\mathcal{M}(\mathcal{C}_1,\mathcal{C}_2, \cdots)$, and the resource consumed, $\mathcal{R}(\mathcal{C}_1,\mathcal{C}_2, \cdots)$; here both the metric and the resources depend on the type of noise, and on a certain set of control parameters, $\{\mathcal{C}_1,\mathcal{C}_2, \cdots\}$. Our aim is to adjust the control parameters to 
minimize the resource under the constraint of both the noise and a desired metric of performance.
One can then define an efficiency of the form
\begin{equation}
\eta(\mathcal{C}_1,\mathcal{C}_2, \cdots) = \frac{\mathcal{M}(\mathcal{C}_1,\mathcal{C}_2, \cdots)}{\mathcal{R}(\mathcal{C}_1,\mathcal{C}_2, \cdots)}.
\label{eq:ETA}
\end{equation}
Then finding the $\{\mathcal{C}_1,
\mathcal{C}_2, \cdots\}$ that maximize this efficiency under the constraint that  $\mathcal{M}(\mathcal{C}_1,\mathcal{C}_2, \cdots)$ takes a given target value, is typically the same as minimizing the resources under that constraint.
As we are taking a hardware agnostic approach in this work, we concentrate on optimizing the VQE algorithm rather than the hardware. Hence, we must consider a resource that makes sense at the level of the algorithm, leading to our choice $\mathcal{R}=\Delta$, where $\Delta$ is explained above.
As already mentioned, here we take $N_\textrm{g}$ and $N_\textrm{it}$ as the two control parameters $\{\mathcal{C}_1,\mathcal{C}_2\}$ that we vary, for a fixed number of qubits.
A natural metric of performance is the algorithm's accuracy, which we can define as $\mathcal{M}= (1- \delta\mathcal{E})$, where $\delta{\cal E}$ is the error in the algorithm's prediction of the quantum system's ground state energy.  Then our goal is to find the $N_\textrm{g}$ and $N_\textrm{it}$ that maximize the efficiency 
(by minimizing $\mathcal{R}$) for a desired algorithm accuracy. This can then be integrated into full-stack models in which one aims to optimize both the algorithm  and the hardware 
\cite{marco2023, trapped_ion_energetics, quandela_BS,qpe_tradeoff} to minimize the physical resources required for a desired precision of the calculation of interest.

The paper is organized in the following manner. Sec.~\ref{sec:VQE_MNR} summarizes the VQE algorithm, using the Heisenberg Hamiltonian as an example with a typical ansatz. It then focuses on establishing scaling relations in the noiseless case in Sec.~\ref{sec:VQE_scaling}, and the noisy case in Sec.~\ref{sec:noisy}. Sec.~\ref{sec:tradeoffs} then explains the use of these scaling relations to characterize efficiency for a VQE algorithm in the presence of noise and infer optimal algorithmic parameters maximizing the efficiency, including in resource-constrained scenario. Sec.~\ref{sec:qpu} shows that our phenomenological scaling relations (developed for errors induced by global-depolarizing noise without error mitigation) also apply for simulations of a realistic noise model (taken from real experimental hardware) combined with error mitigation. Lastly, Sec.~\ref{sec:conclusions} provides our concluding remarks and future outlook.

\section{Scaling relations for Variational Quantum Eigensolver}
\label{sec:VQE_MNR}
The Variational Quantum Eigensolver (VQE) is one of the best-known hybrid classical-quantum algorithms, for a review see Ref.~\cite{Tilly2022}. This algorithm's goal is to find the ground-state energy of the Hamiltonian $\hat{H}$ of a given many-body system, e.g.,\ spin lattice model or molecular system. 
Taking the ground-state energy as a loss function, the VQE algorithm should converge to a reasonable approximation of:
\begin{eqnarray}
    \min_{\{\theta\}} E\big(\{\theta\}\big)
    \ =\  \min_{\{\theta\}} \,\Big\langle \psi \big(\{\theta\}\big)\Big| \,\hat{H} \,\Big| \psi\big(\{\theta\}\big)\Big\rangle  ~,
\end{eqnarray}
where $\{\theta\}$ is the set of variational parameters, and 
$|\psi\big(\{\theta\}\big)\rangle$ is a quantum state synthesized by the quantum circuit in Fig.~\ref{fig:HVA}, e.g., the qubits' state just prior to the measurement.   The classical computer's job  (see Fig.~\ref{fig:basic}) is to use the measurement results at the end of the quantum circuit to modify the choice $\{\theta\}$ to try to iteratively find the optimal $\{\theta\}$ that minimizes $E\big(\{\theta\}\big)$.

\subsection{Barren plateaus and fertile valleys}
It is important to note that in developing our methodology to find the minimum resource cost, we imagine that the algorithm of interest has been designed with the capacity to converge, and not get stuck in a regime of so-called barren plateaus (see \cite{Larocca2024} for a review).
Barren plateaus stems from a curse of dimensionality that plagues variational circuits that explore a large space:  the energy landscape $E(\bm{\theta})$ tends to flatten, on average, as the inverse of the dimension of the explored space (and therefore exponentially with the number of qubits for sufficiently random circuits).
Facing this difficulty, either one reduces the dimension of the explored space (but the ansatz is then likely to be efficiently trainable with a classical method \cite{Cerezo2023,lerchEfficientQuantumenhancedClassical2024}), or one starts the optimization in a so-called \textit{fertile valley} \cite{Holmes2022,Puig2024,Mhiri2025,anselme-martinPreoptimizationQuantumCircuits2026}, which is an area of the parameter space with non-zero gradients, whose size does not shrink exponentially with increasing problem size.
In this work, we assume that such a fertile valley has been identified, and that the algorithm starts somewhere in this valley (a so-called \textit{warm-start} \cite{Puig2024,Mhiri2025}).

With this in mind, we take a concrete model below (VQE for 5-spin Heisenberg chain) that is simple enough that it has converged, without exhibiting barren plateaus \cite{Wiersema2020}, in all cases that we have simulated. 
We expect that this toy-model mimics taking a more complicated problem that exhibits barren plateaus, but then performing a warm-start in that problem's fertile valley.
\begin{figure*}
    \centering
\includegraphics[width=\linewidth, trim=0cm 0cm 0cm 0cm, clip]{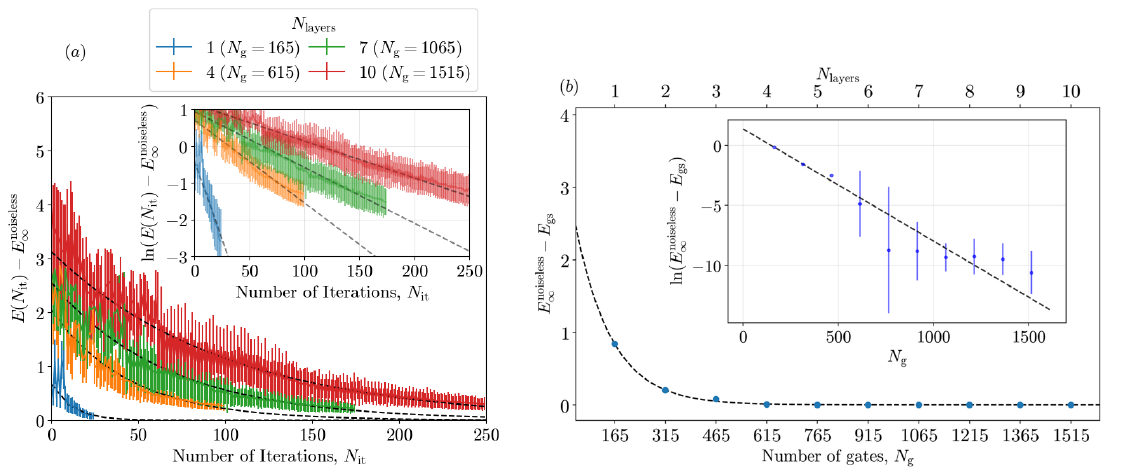}

    \caption{(a) Convergence of the VQE  energy as a function of number of iterations $N_\textrm{it}$, with a varied number of layers ($N_\textrm{layers}$) corresponding to different $N_\textrm{g}$. This convergence can be fit to Eq.~\eqref{eq:noiseless_soln}, and the convergence factor $\mu$ can be extracted using the fit shown as black dashed lines. The same fit is shown in the inset on a log-linear scale. (b) Here we take the plots in (a), along with other similar plots, and extract the converged result of the (noiseless) VQE algorithm ($N_\textrm{it} \to \infty$). We then plot this as a function of the size of the quantum circuit, which is determined by the number of layers in the quantum circuit and quantified by the number of gates in the quantum circuit, $N_\textrm{g}$. The same is shown on the log-linear scale in the inset. This corresponds to the phenomenological relation in Eq.~\eqref{eq:E_ng}. The error bars (too small to see in the main plot, but clearly visible in the inset) represent the standard deviation of the range of values obtained with 50 random seeds, i.e. 50 random initial parameter sets to repeat the algorithm. The datapoints are set at the mean of this range.  The plot shown here is for the noiseless case using the COBYLA optimizer for the circuit shown in Fig.~\ref{fig:HVA}.}
    \label{fig:phen}
\end{figure*}

\subsection{Our toy model: 5-spin Heisenberg chain}
We adopt the Heisenberg spin chain --- ubiquitous in studies of magnetism --- as a toy model to illustrate our findings about the scaling relations governing the relation between accuracy, noise, and algorithmic resource cost. Its Hamiltonian is 
\begin{eqnarray}
    H = J\sum_{i=1}^{n-1} S_i^x S_{i+1}^x + S_i^y S_{i+1}^y + S_i^z S_{i+1}^z~,
    \label{eq:heisen}
\end{eqnarray}
where $i$ is the label for a spin site, and $J$ is the nearest-neighbor coupling constant. We assume that the one-dimensional spin chain is open with a finite number of spins $n$. 
 We work in dimensions where $J=1$, and hence the ground state energy that we aim to calculate with the VQE algorithm will be in units of $J$. For illustrative purposes we consider $n = 5$, for which the ground state energy (found using exact diagonalization) is  
 \begin{eqnarray}E_\textrm{gs} = -7.712~.
\end{eqnarray}
The variational principle guarantees that the energy predicted by the VQE algorithm will never be smaller than this true value of the ground state energy $E_\textrm{gs}$.

The Hamiltonian Variational Ansatz (HVA) \cite{Wiersema2020} is a well-known ansatz for applying the VQE algorithm to the Heisenberg model. 
The ansatz is constructed by decomposing the Hamiltonian, here a spin chain, into distinct non-commuting terms.
One then defines a quantum circuit from this, 
in which one applies $N_\textrm{layers}$ of these non-commuting terms. This circuit is shown in Fig.~\ref{fig:HVA_abstract} for a 5-site chain ($n=5$)
for $N_\textrm{layers}=3$, but we will consider a such circuits with different numbers of layers corresponding to $N_\textrm{layers}$ from 1 to 10. 
Each layer is made of two sub-layers, as shown in Fig.~\ref{fig:HVA_abstract}. The first sublayer is called ``odd (o)'' and corresponds to the interactions between spins $2n$ and $2n+1$, and second sublayer is called ``even (e)'' and corresponding to the interactions between spins $2n$ and $2n-1$.
The  $q$th layer's odd sub-layer has three classical variational parameters ($\theta_{q\textrm{o}}^{(z)}$, $\theta_{q\textrm{o}}^{(y)}$ and $\theta_{q\textrm{o}}^{(x)}$), and  its even sub-layer  has three classical variational parameters ($\theta_{q\textrm{e}}^{(z)}$, $\theta_{q\textrm{e}}^{(y)}$ and $\theta_{q\textrm{e}}^{(x)}$). So that a circuit with $N_\textrm{layers}$ layers is parameterized by $6N_\textrm{layers}$ classical parameters.
The user is free to choose the number of layers in the quantum circuit,  $N_\textrm{layers}$, and our goal here is to find out what is the choice that minimizes the algorithmic resources for given algorithm accuracy.

The number of gates in the quantum circuit, defined as $N_\textrm{g}$, depends on the number of layers, $N_\textrm{layers}$. To find this dependence one must circuit has been transpiled into physical gate operations and count the number of gate operations per layer (plus the number of gate operations in the preparation before the first layer). For this, we consider the transpiled circuit shown in Fig.~\ref{fig:HVA}.  Note that for simplicity, Fig.~\ref{fig:HVA} assumes that all types of gate operations have the same duration, so there is one gate operation on each qubit at each step in the quantum circuit. It is also important to count identity gates (gate-operations in which the qubit's state is simply preserved for a later step of the circuit) in the gate count $N_\textrm{g}$, hence Fig.~\ref{fig:HVA} labels each identity gate with an ``\,\textsf{I}\,''.
We assume that each single qubit gate (including identity gates) is counted once in the gate-count, $N_\textrm{g}$, while each two qubit gate is counted twice. 
We thus assume that if a two-qubit gate acts on a pair of qubits, the error probability and resource cost are the same as if we replaced that two-qubit gate by a pair of single-qubit gate acting on that pair of qubits. This assumption is a simplification, but it is fairly consistent with typical hardware benchmarking.
The transpilation shown in Fig.~\ref{fig:HVA} then has 
150 gates per layer, with an extra 15 gates in the preparation.
After the circuit, a
small subroutine is often appended prior to the  measurement (such as a single gate operation on each qubit prior to its measurement) necessary to  extract given Pauli strings grouped by commutation relations.
This subroutine is hardware dependent, because it depends what measurements are directly available in the hardware (it is different if the hardware can only directly measure the qubit's z-component, or if it can directly measure all components).
For simplicity, we neglect this pre-measurement subroutine here (so it is not shown in Figs.~\ref{fig:HVA_abstract} and Fig.~\ref{fig:HVA}). This allows us to remain
hardware agnostic, while noting that this pre-measurement subroutine has a significantly smaller contribution to $N_\textrm{g}$ than even a single layer; so neglecting the pre-measurement subroutine makes no significant different to our results.
Hence, we have 
\begin{eqnarray}
N_\textrm{g} = 150\, N_\textrm{layers}+ 15~,
\label{eq:N_g}
\end{eqnarray}   
as our relation between number of gates, $N_\textrm{g}$, and the number of layers, $N_\textrm{layers}$ \footnote{If we had included a pre-measurement subroutine consisting of one gate operation per qubit, it would have changed the 15 into 20. This would have no significant effect on our results.}. 

In addition to the primary algorithmic parameters $N_\textrm{g}$ and $N_\textrm{it}$ that we we will vary to optimize efficiency here, there are other algorithmic parameters that we keep fixed. These include
the number of qubits, the number of shots and the number of measurements. For a given problem of interest the number of qubits and the number of measurements are fixed by the structure of the Hamiltonian being studied. The number of shots is assumed to be sufficient to make the measurement sampling error small enough to neglect (much smaller than errors due to circuit noise). Hence, we assume that these parameters are fixed, rather than included among the optimizable parameters.
 
Having defined the circuit, we are now ready to study how 
the algorithm's accuracy depends on $N_\textrm{g}$ and $N_\textrm{it}$ --- first for noiseless circuits and then for noisy circuits.

\subsection{Phenomenological scaling relations for noiseless variational quantum eigensolver}
\label{sec:VQE_scaling}

The goal of this work is to extract phenomenological scaling relations from detailed simulations and use them to optimize resources. We run the VQE algorithm with 50 random initial parameter sets (i.e., random sets of variational angles as in Fig.~\ref{fig:HVA}). From the resulting dataset of energies at each iteration, and its converged value with a large number of iterations, we fit scaling relations to the mean over these 50 seeds.

Looking at the simulations in Fig.~\ref{fig:phen}(a) suggests the scaling relation,
\begin{eqnarray}
E(N_\textrm{it}, N_\textrm{g}) = \alpha\, e^{-\mu(N_\textrm{g}) N_\textrm{it}} + E_{\infty}^\textrm{noiseless}(N_\textrm{g}),
\label{eq:noiseless_soln}
\end{eqnarray}
where $N_\textrm{it}$ denotes the iteration count, and $N_\textrm{g}$ is the total gate count in Eq.~\eqref{eq:N_g}. The three parameters that allow us to fit the dashed black curves to the simulations in Fig.~\ref{fig:phen}(a) are $\alpha$, $\mu(N_\textrm{g})$ and $E_{\infty}^\textrm{noiseless}(N_\textrm{g})$.  Both $\mu(N_\textrm{g})$ and $E_{\infty}^\textrm{noiseless}(N_\textrm{g})$ depend on $N_\textrm{g}$, while $\alpha$ has such a weak $N_\textrm{g}$-dependence that we treat it as $\alpha=2$ for all $N_\textrm{g}$.

\begin{figure}
    \centering
    \includegraphics[width= \linewidth]{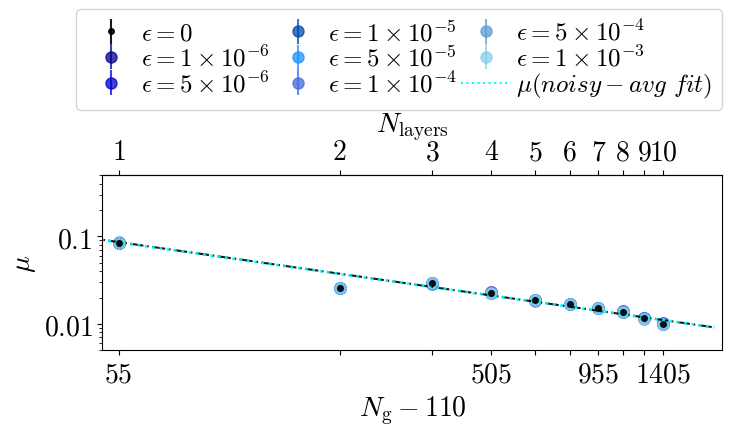}

    \caption{
    The rate at which the VQE energy converges with increasing number of iterations, $\mu$,  plotted versus the number of gates in the circuit, $N_\textrm{g}$. Larger $N_\textrm{g}$  has smaller $\mu$, because larger $N_\textrm{g}$ means more variational parameters to change, and so requires more iterations to converge.
    Here the $\mu$ is defined in Eq.~\eqref{eq:noiseless_soln} for the noiseless case (black), and in Eq.~\eqref{eq:noisy_soln} in the noisy case (various blues). The data for the noiseless VQE algorithm (black dots) in this plot are obtained by extracting $\mu$ from a fitting of simulations of the VQE algorithm (shown in Fig.~\ref{fig:phen}) to Eq.~\eqref{eq:noiseless_soln} and extracting $\mu$ from data with $N_\textrm{it}$ up to about $25 N_\text{layers}$. 
    The  data for the noisy VQE algorithm (blue dots) is extracted in the same way from a fitting  of simulations to Eq.~\eqref{eq:noisy_soln}. From this we see that global depolarizing noise has a negligible effect on the converge rate, $\mu$. 
    We then fit that data with the power relation in Eq.~\eqref{eq:gamma_fit} (dashed lines), where the resulting $\mu_0$, $\lambda$ and  $N_\textrm{g}^\textrm{th}$ are given Table~\ref{table:fit_param_HVA}.
    }
    \label{fig:conv_rate}
\end{figure}

The parameter $E_{\infty}^\textrm{noiseless}(N_\textrm{g})$ represents the converged energy value and approaches the exact ground-state energy $E_\textrm{gs}$ as we increase the number of variational parameters by increasing the number of layers in the circuit (and hence increasing $N_\textrm{g}$), as shown in Fig.~\ref{fig:phen}(b). This trend is captured by the scaling relation, 
\begin{eqnarray}
E_{\infty}^\textrm{noiseless}(N_\textrm{g}) = \beta\, e^{-\kappa N_\textrm{g}} + E_\textrm{gs},
\label{eq:E_ng}
\end{eqnarray}
where $\beta$ and $\kappa$ are fitting parameters governing the convergence behavior.
Fig.~\ref{fig:conv_rate} shows how $\mu$ depends on $N_\textrm{g}$, and we fit this with the power relation,
\begin{eqnarray}
\mu(N_\textrm{g}) = \mu_0\, (N_\textrm{g} - N_\textrm{g}^\textrm{th})^{-\lambda}.
\label{eq:gamma_fit}
\end{eqnarray}
This relation is rooted in the fact that increasing the number of layers in the circuit  introduces more variational parameters, which slows optimization. 

\begingroup
\setlength{\tabcolsep}{8pt} 
\renewcommand{\arraystretch}{1.5} 
\begin{table}
\begin{center}
\begin{tabular}{l  c } 
 \hline
 \hline
 Parameter & Value (arb. units)  \\ [0.5ex] 
 \hline
  Number of random seeds & 50 \\ 
 
 Fitting range of $N_\textrm{it}$ for  E($N_\textrm{it}$)  & 25$N_\textrm{layers}$  \\ 
 
 $\mu_0$ & $1.09   \pm 0.20$  \\
 
  $\lambda$  & $0.63 \pm 0.03$  \\
 
 $N_\textrm{g}^\textrm{th} $ & 110  \\ 
 
 $\beta$ & $3.94$  \\ 
 
 $\alpha$ (fiducial)& 2 \\
 
  $\kappa $ & 0.01  \\ 
 \hline
 \hline
\end{tabular}
\end{center}
 \caption{Values for the phenomenological parameters obtained using fits corresponding to Eqs.~\eqref{eq:noiseless_soln}, \eqref{eq:gamma_fit}, \eqref{eq:E_ng} for the HVA with COBYLA optimizer. A typical value of $\alpha$ is assumed on account of its weak dependence on the number of gates and hence, termed as fiducial.}
 \label{table:fit_param_HVA}
 \end{table}
 \endgroup

The fitted parameters $\mu_0$ and $\lambda$ quantify how the rate of convergence scales with circuit size, whereas $N_\textrm{g}^\textrm{th} $ represents an offset. The three fitted parameters, $\mu_0$, $\lambda$ and $N_\textrm{g}^\textrm{th}$ are found using the standard fitting method (maximizing the coefficient of determination, $R^2$).
For the present circuit and problem instance, the best-fit yields $N_\textrm{g}^\textrm{th} = 110$.

It may be helpful to picture the offset $N_\textrm{g}^\textrm{th}$ as a sort of threshold of circuit complexity; a circuit with less gates than this will have an ansatz that is not sufficient to have any meaningful overlap with the target 
ground state. In such a picture, an ansatz made of the first 22 time-steps in Fig.~\ref{fig:HVA} (for example preparation, the odd sub-layer and first few gates of the even sub-layer) would be about the minimal circuit that gives any meaningful overlap with the target ground-state. However, this is very non-rigorous picture, in reality we treat $N_\textrm{g}^\textrm{th}$ as a phenomenological fitting parameter, so that it is extrapolated from fitting at larger circuit sizes, 
rather than derived from any underlying principles.


We expect that the scaling relations in Eqs.~\eqref{eq:noiseless_soln}--\eqref{eq:gamma_fit} to be typical of many noiseless VQE schemes. 
Table \ref{table:fit_param_HVA} gives the values of the 
the five parameters ($\mu_0$, $\lambda$, $\alpha$, $\beta$, $N_\textrm{g}^\textrm{th}$ and $\kappa$) for the specific VQE scheme simulated here, which we recall is a 5-spin Heisenberg chain using the HVA circuit whose transpilation is shown in Fig.~\ref{fig:HVA}.
However, in general these scaling relations will have values of  the five parameters ($\mu_0$, $\lambda$, $\alpha$, $\beta$, $N_\textrm{g}^\textrm{th}$ and $\kappa$) that depend on (i) the Hamiltonian being studied, (ii) the ansatz used, and (iii) the resulting quantum circuit transpiled into native gates of the hardware.

These scaling relations indicate that the VQE energy in the noiseless regime depends on the algorithmic parameters $N_\textrm{g}$ and $N_\textrm{it}$, and that increasing either of them improves accuracy up to a saturation point, beyond which the gains in accuracy are minimal.
However, this changes when we add the noise that unavoidably occurs in the gate operations, and the scaling relations become more complicated. That is the topic of the next section.

 \subsection{Phenomenological scaling relations for noisy variational quantum eigensolver} 
\label{sec:noisy}

Noise is unavoidable during gate operations, and this profoundly impacts the performance of variational algorithms by causing errors that accumulate with the circuit size. To quantify this effect, we model incoherent noise through a depolarizing channel acting on the circuit state, enabling explicit connection between noise accumulation and algorithmic parameters, particularly the gate count, $N_\textrm{g}$. This model forms the basis for analyzing how noise modifies the VQE algorithm's scaling relations compared to the noiseless case.

Incoherent noise arises at the physical layer of the quantum device and slightly degrades the operation of each gate in the circuit \cite{PhysRevA.106.042421, Fontana2020, Oliv2022, Zeng2020}. 
We model the cumulative effect of such noise by treating it as a global depolarizing channel acting on the density matrix for the quantum state of the qubits, $\rho$ \cite{NielsenChuang2010}:
\begin{eqnarray}
    \mathcal{D} (\rho) = (1-\epsilon)\rho + \epsilon \frac{\mathbb{I}}{2},
    \label{eq:depolarising}
\end{eqnarray}
where $\epsilon$ is the depolarizing probability. While this is about the simplest noise model, it has been shown to provide a good approximation to local noise on each gates in sufficiently random circuits \cite{garcia2024effects, Dalzell2024, Vovrosh_2021} we consider more realistic noise models in Sec.~\ref{sec:qpu}). If there are $\mathcal{N}$ such depolarizing events throughout circuit execution, the evolution of the state obtained from the circuit can be written as
\begin{align}
    \tilde{\rho} = (1-\epsilon)^\mathcal{N}\, U(\theta)\rho U(\theta)^\dagger 
    + \left[1-(1-\epsilon)^\mathcal{N}\right]\, \frac{\mathbb{I}}{2^n},
    \label{eq:depol_state}
\end{align}
where $U(\theta)$ is the noiseless parameterized quantum circuit and $n$ is the number of qubits. This corresponds to a circuit fidelity of $\mathcal{F} = (1-\epsilon)^\mathcal{N}$; more precisely, it corresponds to the circuit outputting the correct 
(noiseless) state with probability $\mathcal{F}$, and outputting an entirely random state with probability $1-\mathcal{F}$.

The expected energy for such a noisy circuit, evaluated with Hamiltonian $H$, is
\begin{eqnarray}
    E_\textrm{noisy}(\boldsymbol{\theta}) &=& [(1-\epsilon)^\mathcal{N}]\, \operatorname{Tr}[H\, U(\boldsymbol{\theta})\rho_0 U(\boldsymbol{\theta})^\dagger] \nonumber\\
    &+& [1-(1-\epsilon)^\mathcal{N}]\, \frac{\operatorname{Tr}(H)}{2^n}.
    \label{eq:rescaled_energy}
\end{eqnarray}
where $\boldsymbol{\theta}$ indicates the set of all classical variational parameters in the circuit. 
For traceless Hamiltonians such as the Heisenberg model in Eq.~\eqref{eq:heisen}, the second term vanishes, effectively scaling the noiseless energy by a factor of $(1-\epsilon)^{\mathcal{N}}$.

The noise-induced degradation of the VQE algorithm's predictions of energy in Eq.~\eqref{eq:rescaled_energy}, can be understood as follows. 
Each time the circuit is run it has a chance $1-(1-\epsilon)^{\mathcal{N}}$ of generating an entirely random state.
However, each iteration of the algorithm requires running that quantum circuit many times (with the same set of variational parameters) to build up the measurement statistics necessary for the classical part of the algorithm to evaluate the quantum state's energy and use that information to choose the variational parameters for the next iteration of the algorithm. 
A proportion $(1-\epsilon)^{\mathcal{N}}$ of these runs will correctly generate the desired noiseless state,  while the others will generate an entirely random state.  
Thus the measured energy given by Eq.~\eqref{eq:rescaled_energy} is simply the weighted average of the energy of the desired (noiseless) state, given by $\operatorname{Tr}[H\, U(\boldsymbol{\theta})\rho U(\boldsymbol{\theta})^\dagger]$ and the average energy of an entirely random state of the Hamiltonian given by $\operatorname{Tr}(H)\big/2^n$ (where our Heisenberg Hamiltonian in Eq.~\eqref{eq:heisen} has $\operatorname{Tr}(H)=0$).

\begin{figure}
    \centering
    \includegraphics[width= \columnwidth]{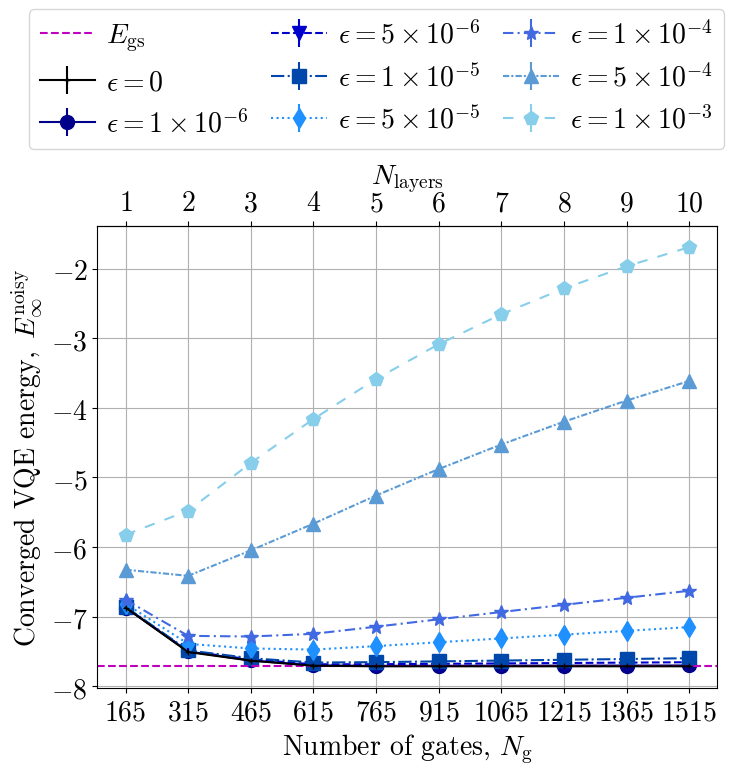}
    \caption{Plot of the converged prediction for the ground state energy ($E^\textrm{noisy}_\infty$ from Eq.~\eqref{eq:E_conv_noisy}) in the presence of noise as a function of $N_\textrm{g}$. The magenta dashed line is the true value of the ground-state energy, $E_\textrm{gs}$. In the noiseless case ($\epsilon = 0$, black), the VQE algorithm's prediction for the ground-state energy (converged in the limit of a large number of iterations) saturates at its most accurate value as $N_\textrm{g} \to \infty$ . However, when noise is present, then increasing the number of gates $N_\textrm{g}$, causes increasingly deviations from the noiseless value, leading to the a U-shaped curve
    with optimum at $N_\textrm{g}^\textrm{opt}(\epsilon)$, corresponds to the VQE algorithm's most accurate prediction of  $E_\textrm{gs}$.
    As in the noiseless case, all the values are obtained using $50$ initial random seeds for each value of $N_\textrm{g}$ and $\epsilon$. We also calculate error bars on each data point by taking the standard deviation over these random seeds, but the error-bars are then
    smaller than the data-point markers and so are invisible.}
    \label{fig:E_ng_tradeoff}
\end{figure}

We assume that all gates operations (including identity gates) have the same noise per qubit, so the number of depolarizing events is simply given by
\begin{eqnarray}
    \mathcal{N} = N_\textrm{g}.
\end{eqnarray}
This makes the link between the circuit's physical noise and the algorithmic parameter $N_\textrm{g}$. 
Substituting this into Eq.~\eqref{eq:E_ng}, the converged energy becomes
\begin{equation}
    E_{\infty}^\textrm{noisy}(N_\textrm{g}) = (1-\epsilon)^{N_\textrm{g}}E_{\infty}^\textrm{noiseless}(N_\textrm{g}).
    \label{eq:E_conv_noisy}
\end{equation}
This captures the competing effects of increased circuit size and noise, as shown for different values of $\epsilon$ in Fig.~\ref{fig:E_ng_tradeoff}. The figure makes it  clear that increasing the number of gates, $N_\textrm{g}$, initially improves accuracy as the circuit has more layers with more variational parameters, and so can span a larger portion of the Hilbert space. However, beyond a certain $N_\textrm{g}$ the depolarization noise starts to dominate, driving the algorithm's prediction away from the correct value. This behavior means that there is an  optimal gate count for any give noise-strength $N_\textrm{g}^\textrm{opt}(\epsilon)$; it gives the circuit size that maximizes the algorithm's accuracy.

\begin{figure*}
    \centering
    \includegraphics[width = \linewidth, trim = 0cm 0.5cm 0cm 0cm, clip]{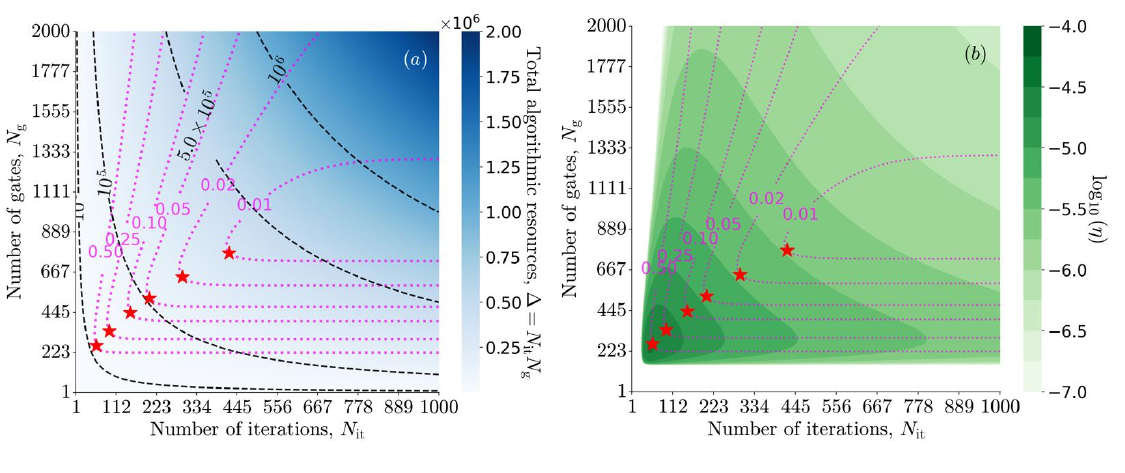}
    \caption{Algorithmic accuracy and algorithmic resources as a function of the parameters $N_\textrm{it}$ and $N_\textrm{g}$, for noise parameter  $\epsilon =10^{-6}$. Magenta dashed curves are contours of fixed algorithm accuracy, defined by iso-error contours, where the error is defined as 
    $\delta \mathcal{E}$ in Eq.~\eqref{eq:error}. 
    In (a) we superimpose these iso-error contours on a
    heatmap of algorithmic resource cost $\Delta = N_\textrm{it} N_\textrm{g}$, while in (b) we superimpose them on a heatmap of the efficiency $\eta$ in Eq.~\eqref{eq:efficiency}). A darker shade of blue denotes larger value of $\Delta$ in (a), whereas a darker shade of green represents higher $\eta$ in (b). The iso-resource cost: $\Delta$ lines are shown as black-dashed in (a). The excluded region in (b), here shown as white is where the error exceeds 1, and hence the metric is negative, deeming it impossible to be plotted on the logarithmic scale of the heatmap colorbar. The red stars $\red{\star}$ represent the minimum $\Delta$ for each iso-error contour in both (a) and (b). }
    \label{fig:energetics}
\end{figure*}


Having established how noise affects the converged accuracy ($N_\textrm{it} \to \infty$) through Eq.~\eqref{eq:E_conv_noisy}, we now turn to a joint description that explicitly involves both the number of iterations, $N_\textrm{it}$ and the gate count, $N_\textrm{g}$. 
The rate of convergence with increasing  $N_\textrm{it}$ is given by  $\mu$ in Eq.~\eqref{eq:gamma_fit}). In the noiseless setting, $\mu$ decreases with increasing $N_\textrm{g}$, reflecting the slower optimization typically associated with circuits with more variational parameters. The question is whether noise modifies this relationship by distorting the optimization landscape or altering the effective parameter dynamics.

For global depolarizing noise considered here, given by Eq.~\eqref{eq:depolarising}, the answer is no. From Eq.~\eqref{eq:rescaled_energy}, we can see that the noise acts as a uniform rescaling of the circuit's prediction of the energy (see the discussion below Eq.~\eqref{eq:rescaled_energy}).
If we define the energy landscape as the energy  $E_\textrm{noisy}(\boldsymbol{\theta})$ as a function of the set of variational parameters given by $\boldsymbol{\theta}$, then we see that the noise  
does not deformation of the energy landscape, it is only causes a uniform suppression of the amplitude of the energy landscape by a factor of $(1-\epsilon)^{N_\textrm{g}}$. 
Now, the way the algorithm explores the energy landscape (for example using a steepest descend method) depends on that landscape's shape not its amplitude, so the optimizer explores the noisy energy landscape in the same way that it explores the noiseless landscape. Thus we expect $\mu$ to exhibit no dependence on noise strength $\epsilon$, as confirmed by Fig.~\ref{fig:conv_rate}.

Bringing the above information together, the accuracy of the noisy VQE algorithm's prediction is a function two parameters; the number of iterations, $N_\textrm{it}$ and the gate count, $N_\textrm{g}$. For the global depolarizing noise, this is given by  Eqs.~\eqref{eq:noiseless_soln}, \eqref{eq:E_ng}, and \eqref{eq:rescaled_energy}, resulting in
\begin{eqnarray}
E^\textrm{noisy} (N_\textrm{it}, N_\textrm{g}) = \big[(1-\epsilon)^{N_\textrm{g}}\big] \alpha e^{-\mu(N_\textrm{g})N_\textrm{it}} 
+ E_{\infty}^\textrm{noisy}  (N_\textrm{g})~. \nonumber\\
\label{eq:noisy_soln}
\end{eqnarray}
Combining this with Eq.~\eqref{eq:E_conv_noisy}, and the  phenomenological scaling relation in Eq.~\eqref{eq:gamma_fit} gives
\begin{eqnarray}
    E (N_\textrm{it}, N_\textrm{g}) &=&  (1-\epsilon)^{N_\textrm{g}}\Big[ \alpha e^{-\mu_0 N_\textrm{it} (N_\textrm{g}-N_\textrm{g}^\textrm{th})^{-\lambda}} \nonumber \\
    & & \qquad \qquad \qquad + \beta e^{-\kappa N_\textrm{g}} + E_\textrm{gs} \Big]\,. \ \ 
    \label{eq:gen_soln}
\end{eqnarray}

To assess whether the scaling relations are specific to the HVA or reflect more general features of variational optimization, we repeat the analysis using the RY ansatz (RYA)~\cite{Bravo-Prieto2020}, a hardware-efficient ansatz with no direct correspondence to the target Hamiltonian. As shown in Appendix~\ref{app:RYA}, all three functional forms --- Eqs.~\eqref{eq:noiseless_soln}, \eqref{eq:E_ng}, and~\eqref{eq:gamma_fit} --- persist with different numerical coefficients (Table~\ref{table:fit_param_RYA}), supporting the generality of the phenomenological scaling relations.

Our next step is to return to the Hamiltonian Variational Ansatz (HVA), and explore the subtle interplay of noise, accuracy and algorithmic resources represented by Eq.~(\ref{eq:gen_soln})
with the parameters listed in Table~\ref{table:fit_param_HVA}.

\begin{figure}[t]
\centering
\includegraphics[width=\columnwidth]{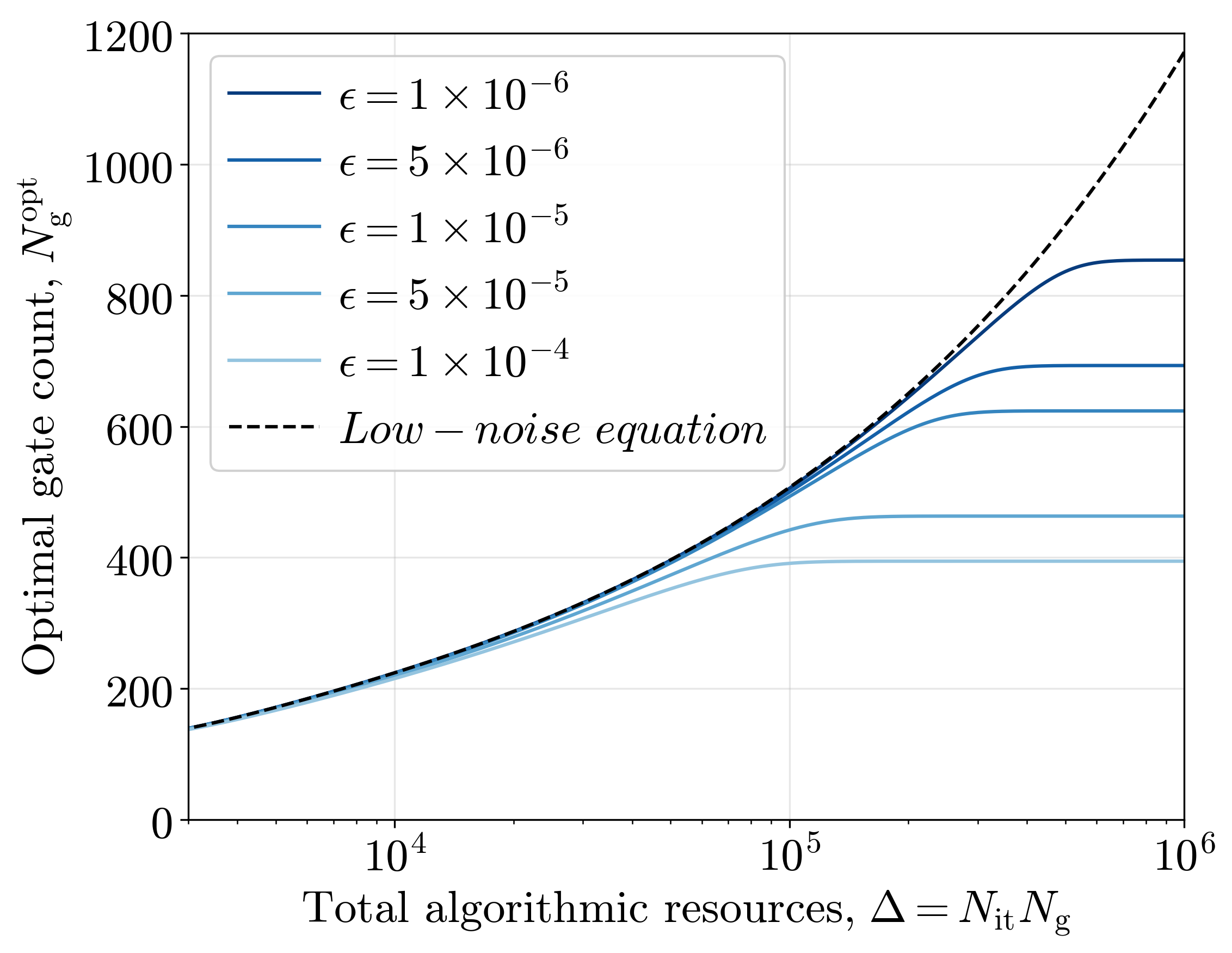}
\caption{Optimal gate count $N_\textrm{g}^\textrm{opt}$ as a function of total algorithmic resource cost $\Delta$. Solid lines: numerical solutions to the full stationarity condition Eq.~\eqref{eq:allsubequations-exact_stationarity} for various noise rates $\epsilon \in [10^{-6}, 10^{-4}]$. Dashed black line: low-noise solution from Eq.~\eqref{eq:low_noise_balance}. The saturation of $N_\textrm{g}^\textrm{opt}$ at large $\Delta$ indicates a noise-dominated regime where additional resources yield no benefit; see Eq.~\eqref{eq:stationarity_largeDelta_exact-3} for the analytical limit.}
\label{fig:Ng_vs_Delta}
\end{figure}
\begin{figure}[t]
    \centering
    \includegraphics[scale = 0.45]{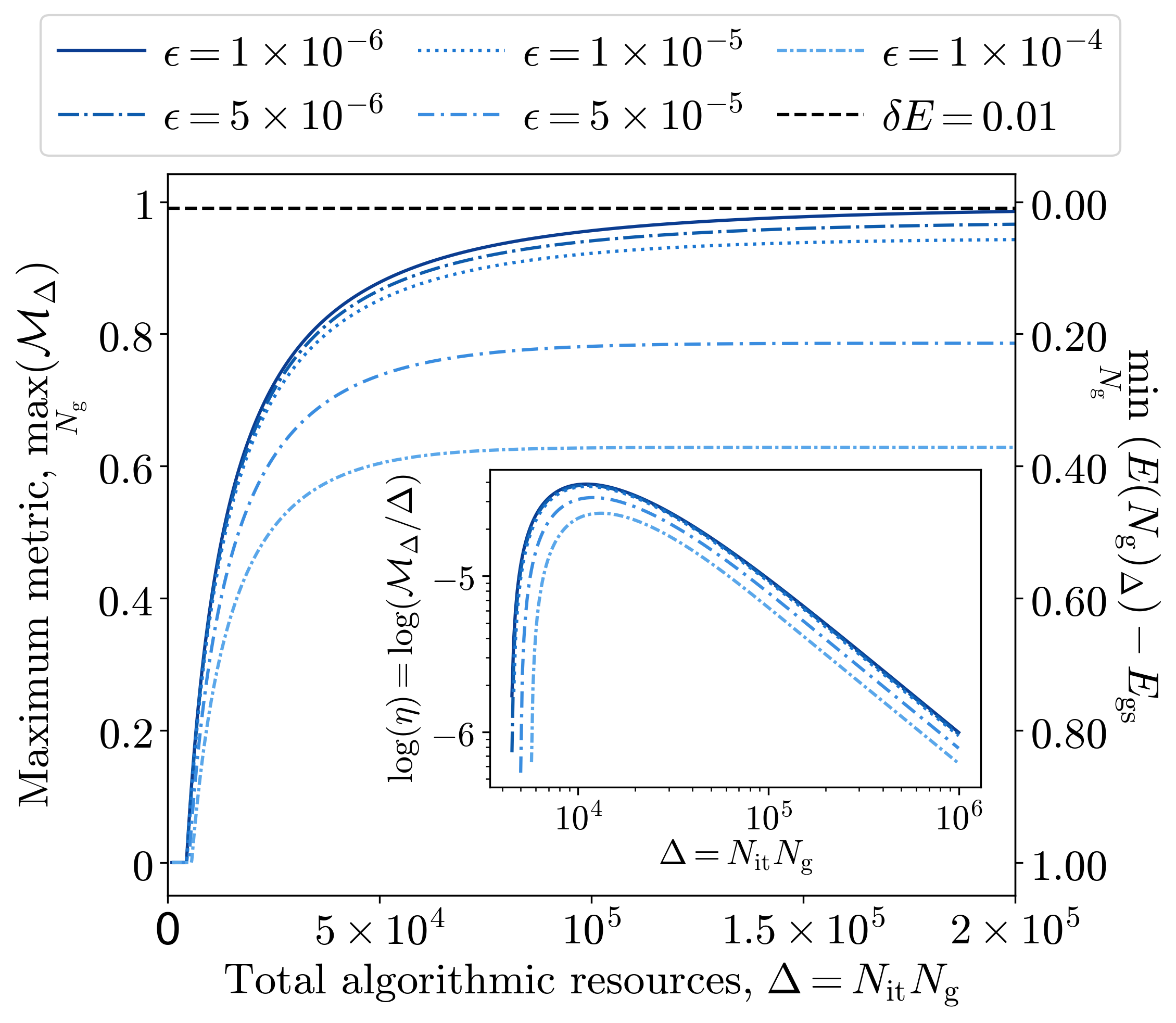}
            \caption{Maximum metric obtained from Eq.~\eqref{eq:E_fixed_metric} on the left vertical axis (y-axis) and the corresponding minimum error on the right vertical axis (with respect to the fixed algorithmic resource $\Delta = N_\textrm{it} N_\textrm{g}$) as a function of the total algorithmic resource. As expected, increasing the amount of resource leads to a lower error in the  energy, and saturation to a value determined by $\epsilon$. 
            For higher values of $\epsilon$, the maximum metric remains low at all $\Delta$, so the algorithm gives poor results even if one uses a lot of resources (i.e., even if $\Delta\to\infty$). 
            The corresponding efficiency from Eq.~\eqref{eq:efficiency} is shown in the inset, as a function of the algorithmic resource.}
    \label{fig:best_metric}
\end{figure}

\section{Resource optimization}
\label{sec:tradeoffs}

We employ the phenomenological scaling relations derived in Sec.~\eqref{sec:noisy} to analyze and minimize resource consumption in the presence of noise, as well as to identify operating points of maximal efficiency under a fixed algorithmic resource cost. Treating the number of iterations, $N_{\textrm{it}}$, and the per-iteration gate count, $N_\textrm{g}$, as the primary algorithmic parameters, the algorithmic resource cost is taken to be the total number of gate operations in the algorithm, $\Delta = N_{\textrm{it}} N_\textrm{g}$. This $\Delta$ is the quantity that we want to minimize, since that will typically minimize both total runtime and energy consumption of the algorithm, and we want to minimize this while preserving algorithmic accuracy.

To quantify the algorithmic accuracy, we define algorithm's error
as 
\begin{eqnarray}
        \delta \mathcal{E}  &=&  \frac{E^\textrm{noisy}(N_\textrm{it}, N_\textrm{g})-E_\textrm{gs}}{J}\, .
        \label{eq:error}
    \end{eqnarray}
where $J$ is the typical energy scale of the Hamiltonian of interest (see Eq.~\ref{eq:heisen}), but without loss of generality we set $J=1$.
Then our metric of success of the algorithm is $1- \delta \mathcal{E}$, so this metric grows as the algorithm gets closer to the true ground-state energy, and a metric of 1 corresponds to a algorithm that perfectly finds the ground-state energy.
The resource efficiency is then defined as in Eq.~\eqref{eq:ETA},
giving
    \begin{eqnarray}
        \eta &=& \frac{1-\delta\mathcal{E}}{\Delta} \,,
        \label{eq:efficiency}
    \end{eqnarray}
This efficiency grows monotonically as the algorithm becomes more precise (smaller $\delta \mathcal{E}$) or the algorithmic resource cost $\Delta$ is reduced.

 Fig.~\ref{fig:energetics}(a) plots the total resource heatmap, and Fig.~\ref{fig:energetics}(b) plots the efficiency heatmap. 
 Both plots also show the iso-error contours, where we have fixed the noise parameter to be $\epsilon = 10^{-6}$.
 For this, these plots use the scaling relation in Eq.~\eqref{eq:gen_soln} with the phenomenological fit parameters listed in Table~\ref{table:fit_param_HVA}.  The efficiency heatmap has an exclusion region (white) corresponding to $\delta\mathcal{E} > 1$, due to the logarithmic scale. Both of these heatmaps contain points of maximal efficiency, or equivalently minimum $\Delta$, for each iso-metric curve; marked as red stars ($\red{\star}$) on the curves. 
It is important to highlight that in Fig.~\ref{fig:energetics}(a) the total algorithmic resources for given metric can be reduced by significant factors by operating an the red star, rather than somewhere else on the iso-error contour. For instance, for $\delta\mathcal{E} = 0.01$, the optimal point consumes less than half the resources of elsewhere in the plot, while for $\delta\mathcal{E} = 0.05$ the optimal point consumes nearly 10 times less resources than elsewhere in the plot.
Hence, working at the optimal point will deliver both a significant gain in both the speed of the algorithm, and a significant reduction in energy consumption.

In summary, Eq.~\eqref{eq:gen_soln} provides a minimal yet complete description, linking the algorithmic resources $(N_\textrm{g}, N_{\textrm{it}})$, the noise parameter $\epsilon$, and the resulting accuracy. This functional form enables the analytic extraction of optimal gate counts that ensure maximum efficiency under the constraint of given (desired) algorithm accuracy. It also gives the maximum accuracy under resource constraints, should the total number of gate operations be constrained by available runtime or other considerations.

\subsection{Resource-constrained analysis} 
\label{sec:resources}


Since we want to find the number of gate-operations that minimizes the algorithm resources, $\Delta$,  for given algorithm precision,
let us fix $\Delta$, and replace $N_\textrm{it}$ by $\Delta/N_\textrm{g}$ in Eq.~\eqref{eq:gen_soln}. This gives 
\begin{eqnarray}
    E^\textrm{noisy}_{\Delta} (N_\textrm{g}) &=& (1-\epsilon)^{N_\textrm{g}}
    \Big(\alpha \,e^{-\mu_0 \Delta \,f(N_\textrm{g})} 
    + \beta e^{-\kappa N_\textrm{g}} + E_\textrm{gs}\Big),\nonumber\\
    \label{eq:E_fixed_metric}
\end{eqnarray}
where $f(N_\textrm{g}) \equiv 1\big/(N_\textrm{g}(N_\textrm{g}-N_\textrm{g}^\textrm{th})^\lambda\big)$. This expression gives the VQE energy as a function of circuit size alone, with the iteration count implicitly determined by the constraint $N_\textrm{it} = \Delta / N_\textrm{g}$. Geometrically, varying $N_\textrm{g}$ at fixed $\Delta$ corresponds to moving along the iso-resource lines (black-dashed) in Fig.~\ref{fig:energetics}(a).

The form of Eq.~\eqref{eq:E_fixed_metric}, admits a unique minimum at $N_\textrm{g}^\textrm{opt}$ corresponding to red stars in  Fig.~\ref{fig:energetics}(a).
This means that $N_\textrm{g}=N_\textrm{g}^\textrm{opt}$ minimizes the algorithmic resources for a given (desired) algorithm accuracy, and it also maximizes this accuracy for fixed resource consumption. 

 Eq.~\eqref{eq:E_fixed_metric} is determined by the competition between three effects: convergence rate of the optimizer (smaller $N_\textrm{g}$ converges faster, allowing smaller $N_\textrm{it}$), ability of the ansatz to converge to a accuracy result (favoring large $N_\textrm{g}$), and noise accumulation (penalizing large $N_\textrm{g}$). The optimal gate count is that which maximizes the algorithm's accuracy for given $\Delta$ (by minimizing $E^\textrm{noisy}_\Delta$), and so it is given by the stationarity condition $\partial E^\textrm{noisy}_\Delta / \partial N_\textrm{g} = 0$, yielding a transcendental equation derived in Appendix~\ref{app:Ngopt}.
In the low-noise limit ($\epsilon \to 0$) with $N_\textrm{g} \gg N_\textrm{g}^\textrm{th}$, the stationarity condition for optimal $N_\textrm{g}$ simplifies to:
\begin{eqnarray}
\alpha(1+\lambda)\mu_0 \Delta \exp\left[-\frac{\mu_0\Delta}{N_\textrm{g}^{1+\lambda}}\right] \simeq \beta \kappa N_\textrm{g}^{2+\lambda} e^{-\kappa N_\textrm{g}}.
\label{eq:low_noise_balance}
\end{eqnarray}
Fig.~\ref{fig:Ng_vs_Delta} compares the numerical solution of the full stationarity condition (solid lines, various $\epsilon$) with the low-noise prediction in Eq.~\eqref{eq:low_noise_balance} (dashed black line). For $\epsilon \leq 10^{-4}$, the curves closely follow this low-noise prediction. At larger $\Delta$, finite-$\epsilon$ curves deviate and saturate as noise accumulation becomes significant.

Fig.~\ref{fig:best_metric} shows the best achievable metric $\max_{N_\textrm{g}}(\mathcal{M}_\Delta)$ as a function of the total algorithmic resource cost. The metric improves with $\Delta$ before saturating at a noise-dependent ceiling: higher $\epsilon$ leads to earlier saturation at larger error. Achieving practical target accuracies (e.g., $\delta\mathcal{E} = 0.01$) requires substantial resources and sufficiently low noise. 
This optimal relation between metric of success (algorithm accuracy) and the algorithmic resource consumption,  $\Delta$, can be read either as the maximum metric of success for given algorithmic resource, or the minimal algorithmic resource for given (desired) metric of success.

Fig.~\ref{fig:best_metric}'s inset shows the corresponding efficiency $\eta = \mathcal{M}/\Delta$, which initially increases with $\Delta$ as accuracy improves rapidly, then decreases as the metric saturates while resources continue to grow. 

The fact that this efficiency is peaked is an indication that 
as one increases the accuracy, then further accuracy increases become increasingly costly in resources. However, only the end-user of the algorithm knows what accuracy they require. Thus, end-users will want to see the whole curve of accuracy versus resources to establish the minimum resource consumption for their desired accuracy, or to maximize the accuracy under an given constraint on the resources.

In summary, the resource-constrained analysis identifies an optimal depth-iteration allocation that maximizes accuracy for a given algorithmic resource cost. Operating near $N_\textrm{g}^\textrm{opt}$ ensures efficient utilization of limited resources. This provides a principled criterion for circuit depth selection in a noisy VQE algorithm, enabling direct comparison of which ansatz achieves a target accuracy with minimal resource cost.

The utility of this optimization method depends on whether our scaling relations survive in the realm of real quantum processors, where noise will be more complicated (not the global depolarizing noise assumed above), and the algorithms will be run using error mitigation techniques.  This is what we address in the next section.

\begin{figure*}
    \centering
    \includegraphics[width=\linewidth]{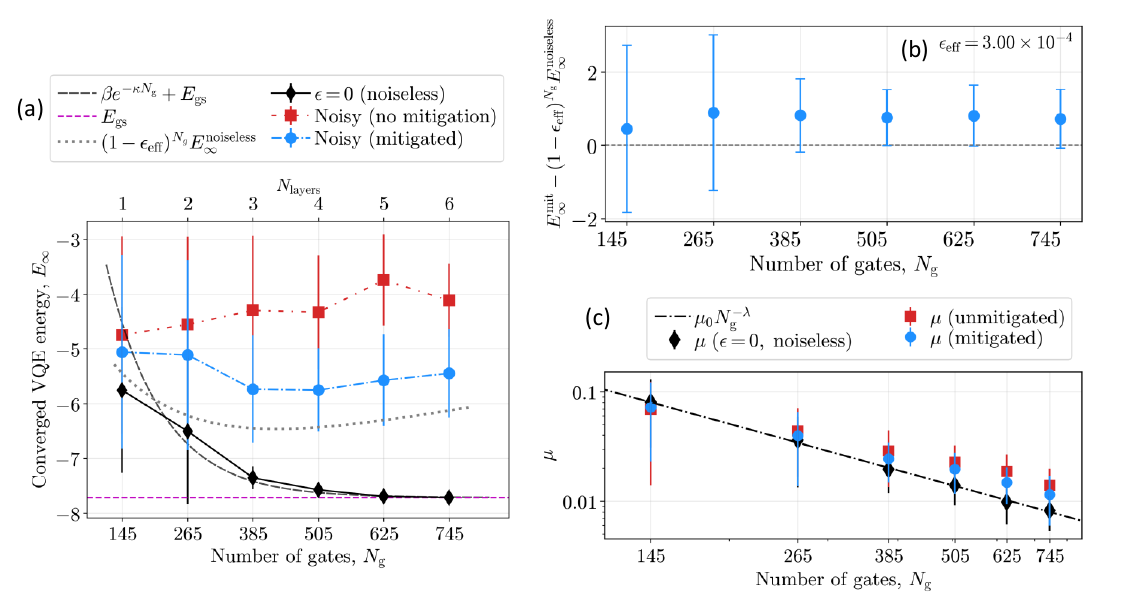}
    \caption{Validation of our phenomenological scaling relations on a noisy IBM Nighthawk fake backend, with and without error mitigation, for the 5-qubit open-chain Heisenberg model using the HVA ansatz.
    (a) Converged VQE energy $E_\infty$ as a function of the total gate count $N_\textrm{g}$ (equivalently, the number of HVA layers $N_{\rm layers}$, top axis). Black diamonds: noiseless reference ($\epsilon = 0$), with the dashed black curve showing the fit $\beta\,e^{-\kappa N_\textrm{g}} + E_{\rm gs}$ from Eq.~\eqref{eq:E_ng}. Red squares are the noisy hardware without mitigation; this is clearly in the noise-dominated regime where the algorithm does not work, because it never converges (so there is no point in resource optimization). Blue circles: noisy hardware with error mitigation denoted as $E_\infty^\textrm{mit}$. The error-mitigated data indicate a tradeoff --- a slight increase in accuracy followed by a decrease --- with $N_\textrm{g}$. The gray dotted curve shows the global-depolarizing prediction $E_\infty^\textrm{fit} = (1-\epsilon_{\rm eff})^{N_\textrm{g}}\,E_\infty^{\rm noiseless}$ with $\epsilon_{\rm eff} = 3.00 \times 10^{-4}$.
    (b) Residuals of the mitigated converged energies relative to the fitted curve for the global-depolarizing model, $E_\infty^{\rm mit} - E_\infty^{\rm fit}$ showing agreement within error bars across the full range of $N_\textrm{g}$, but with appreciable scatter, confirming that effective noise after error mitigation is only approximately global-depolarizing. (c) Convergence rate $\mu$ versus $N_\textrm{g}$, comparing noiseless values (black diamonds) with those extracted from mitigated runs (blue circles). Both follow the power relation $\mu_0\,N_\textrm{g}^{-\lambda}$ (dash-dotted line) within error bars. The red squares in this panel denote the converge rate obtained from unmitigated data. In the panels, 20 random seeds are used for the initial variational parameters; markers indicate the mean,  while the error bars indicate the standard deviation over seeds. The COBYLA optimizer is used throughout.}
    \label{fig:noisy_baseline_qpu}
\end{figure*}

\section{Phenomenological scaling relations on real quantum processors}
\label{sec:qpu}

In previous sections we developed  phenomenological scaling relations using a simple model of noise (global depolarizing noise) with no error mitigation.
A natural question is whether these scaling relations work when the VQE algorithm is executed on real quantum processing unit (QPU), which exhibits much more complicated noise, and which uses error mitigation to reduce (but not eliminate) the impact of this noise. To be more specific, real hardware typically experiences local noise on each qubit, which may vary significantly from qubit to qubit, and exhibit crosstalk between qubits. This is much more complicated than the simple noise model treated above. In addition, most such hardware uses at least one error mitigation technique, such as ``zero-noise extrapolation via gate-folding'' (see Appendix~\ref{sec:mitigation}).

The noiseless relations in Eqs.~\eqref{eq:noiseless_soln} and~\eqref{eq:E_ng} are, by construction, insensitive to hardware-specific noise and error mitigation. Nevertheless, we re-establish the noiseless scaling relations based on the abstract circuit in Fig.~\ref{fig:HVA_abstract} with a noiseless emulation based on IBM \textit{Nighthawk} QPU characteristics such as connectivity and gateset. The pass manager, which handles transpilation and compilation is free to perform hardware specific conversion of the circuit. 
We observe that for this emulation, while the functional forms of Eqs.~\eqref{eq:noiseless_soln} and~\eqref{eq:E_ng} remain the same, a change in the values of phenomenological parameters such as $\beta, \kappa$ is expected. The noiseless converged energies as a function of the number of gates are shown in Fig.~\ref{fig:noisy_baseline_qpu} (a) as black diamonds. The dashed gray line denotes a fit of Eq~\eqref{eq:E_ng}. Similarly, the convergence rate $\mu$ in the noiseless case is shown as black diamonds in Fig.~\ref{fig:noisy_baseline_qpu} (c). The extracted convergence rate follows the power relation $\mu = \mu_0 N_\textrm{g}^{-\lambda}$. These fits for the noiseless case fix the reference values of $\beta$, $\kappa$, $\mu_0$, and $\lambda$.

What remains to be studied is whether two noise-dependent predictions survive on a real QPU: (i) the noise bias in the converged energy (Eq.~\eqref{eq:E_conv_noisy}) and (ii) the convergence rate with iterations (Eq.~\eqref{eq:gamma_fit}). This is not guaranteed as noise on a real device generally deviates from the global depolarizing model assumed in Eq.~\eqref{eq:rescaled_energy}. To test these, we run the abstract circuit in Fig.~\ref{fig:HVA_abstract} on an IBM Qiskit~\cite{qiskit} backend simulator configured with a \textit{fake QPU} that mimics the noise characteristics of a real QPU, with different noise on each qubit extracted from experimental tests on the real hardware. Appendix~\ref{sec:sup_qpu} provides the details of this implementation --- selection of QPU and qubits, etc. 

From this implementation, in the absence of error mitigation, we obtain the noisy converged VQE energy given as the red squares in Fig.~\ref{fig:noisy_baseline_qpu}(a). Even with compiler-level optimization enabled, the converged energy sits in the \textit{noise-dominated regime}: the usual competition between accuracy gain from deeper circuits and bias accumulation from noise is invisible, and accuracy instead degrades monotonically with $N_\textrm{g}$. The red squares (corresponding to noisy runs) in Fig.~\ref{fig:noisy_baseline_qpu} lie several units of energy above the noiseless reference (black diamonds) and show no improvement as $N_\textrm{layers}$ increases from $1$ to $6$. Thus is the absence of error mitigation, the algorithm fails to work, and so no meaningful optimization over algorithmic resources is possible. 

Thus, we implement the error mitigation technique known as ``zero-noise extrapolation via gate-folding''  (as discussed in Appendix~\ref{sec:mitigation}) on this simulator of real QPU hardware.
The error-mitigated converged energies ($E_\infty^\textrm{mit}$) are shown as blue circles in Fig.~\ref{fig:noisy_baseline_qpu}(a) for which we observe a weak dependence on circuit-size, $N_\textrm{g}$. 
We see that behavior is the same as predicted by our phenomenological scaling relation; as we increase $N_\textrm{g}$, the accuracy initially improves and then getting worse, with highest accuracy for a circuit with 3 or 4 layers ($N_\textrm{g}$=385 or 505).
Hence, we fit the rescaled converged energy with Eq.~\eqref{eq:E_conv_noisy}, and show $ E_\infty^\textrm{fit} = (1-\epsilon_\textrm{eff})^{N_\textrm{g}}\, E_\infty^\textrm{noiseless}$ as a gray dotted line in Fig.~\ref{fig:noisy_baseline_qpu} (a). We obtain the effective depolarization parameter, $\epsilon_\textrm{eff} = 3 \times 10^{-4}$, using noiseless parameters such as $\kappa$ and $\beta$. The residuals $E_\infty^\textrm{mit} - E_\infty^\textrm{fit}$ are shown for each value of $N_\textrm{g}$ in Fig.~\ref{fig:noisy_baseline_qpu}(b). They are approximately within error bars of zero across the full range of $N_\textrm{g}$, but the scatter is appreciable, suggesting that error-mitigated hardware noise is only approximately global-depolarizing. 
Fig.~\ref{fig:noisy_baseline_qpu}(c) shows that the error-mitigated convergence rates $\mu_\textrm{mit}$ overlap with the noiseless power relation within their (large) error bars, indicating that this error-mitigation preserves the exponent $\lambda$ governing how the convergence rate changes with circuit depth. All this suggests that the error-mitigated results (for the realistic hardware) are different from global depolarizing noise model, but close enough that the model is a reasonable approximation. 
This is sufficient to use our scaling relations to give reasonable predictions of the parameters  ($N_\textrm{it}$ and $N_\textrm{g}$) that give the best accuracy for given algorithmic resource cost, for minimize this resource cost for desired accuracy.

At this point we emphasize that, while our scaling relations were developed (and tested) using data from the noiseless algorithm, we do not need any such noiseless data to find the phenomenological parameters in the scaling relations ($\alpha$, $\beta$, $\mu_0$, $\lambda$, $\kappa$, and 
$N_\textrm{g}^{\textrm{th}}$).
This is crucial 
when working with real quantum hardware (with circuits that are too large to simulate on classical computers), because one can never turn off the noise in real quantum hardware.  
To this end, Appendix~\ref{app:outlook} proposes a self-contained resource optimization pipeline which uses data from noisy hardware {\it alone} to estimate the noise strength, find the phenomenological parameters in our scaling relations ($\alpha$, $\beta$, $\mu_0$, $\lambda$, $\kappa$, and 
$N_\textrm{g}^{\textrm{th}}$), and then optimize algorithmic resources.

\section{Conclusions and future outlook}
\label{sec:conclusions}

We presented a systematic analysis of Variational Quantum Eigensolver (VQE) algorithm's resource efficiency using the 
one-dimensional spin-$\frac{1}{2}$ isotropic Heisenberg model as a representative 
benchmark. Within the Metric-Noise-Resource (MNR) framework, we introduced 
phenomenological scaling relations that capture the interplay of algorithmic accuracy, the number of gates in the circuit $N_\textrm{g}$, the number of iterations of that circuit 
$N_{\textrm{it}}$, noise strength $\epsilon$, and the total algorithmic resource cost (quantified by 
$\Delta = N_\textrm{g} N_{\textrm{it}}$).

From numerical simulations, we extracted scaling relations revealing three 
central features: (i) In the absence of noise, larger circuits give more accurate results, although the improvement in accuracy saturates when the circuit becomes very large.
(ii) In the absence of noise, there is a systematic slowdown of convergence for larger circuits.
(iii) In the presence of noise, larger circuits have more errors; this means that increasing the circuit size makes the algorithm more accurate while errors are rare, but this accuracy degrades when one further increases the circuit size into the regime where errors become significant. 
The resulting structure parallels scaling-relation behavior 
in classical machine learning, suggesting that noisy VQE algorithm obeys similarly 
structured depth-optimization trade-offs.

Our scaling relations contain various phenomenological parameters ($\alpha$, $\beta$, $\mu_0$, $\lambda$, $\kappa$, and 
$N_\textrm{g}^{\textrm{th}}$), which will depend on the ansatz, the
Hamiltonian, the optimizer, and the noise model. They can be extracted empirically from a small number of runs of the quantum circuit (as we demonstrated here for both the Hamiltonian Variational Ansatz (HVA) and the RY Ansatz).

Using these scaling relations, we 
obtained 
closed-form solutions for the optimal number of gate-operations in the quantum circuit, which 
maximized the algorithm's accuracy for a given algorithmic resource cost. This optimal operating point is given by a transcendental equations --- solvable numerically --- that avoids needing exhaustive parameter sweeps. 

We demonstrate that our scaling relations will work for real experimental hardware, which typically exhibits different noise on each qubit,  and require error mitigation. This is done by executed the VQE algorithm on a noisy backend emulating the IBM Nighthawk processor
(which emulates realistic noise on each qubit), with a commonly used error-mitigation technique. 

Future directions would be to test our phenomenological scaling relations on different NISQ algorithms, larger numbers of qubits, other types of noise, and other types of error mitigation. 
At the same time, it would be very useful to extend the scaling relations to optimizing other parameters that impact both algorithm accuracy and algorithmic resource costs, such as the amount of error-mitigation or the number of shots used to get measurement statistics.
In any case, optimizing the speed and energy consumption of near-term quantum computers requires full-stack models (which allow the joint optimization of hardware and algorithms), and such scaling relations for the algorithmic part will be crucial in the construction of such full-stack models.

\begin{acknowledgments} 
The authors thank Martin Plazanet, Kong Jian Feng, Wong Zi Cheng, Lorenzo Buffoni, Alessandro Luongo, Olivier Ezratty, Christophe Domain, Adrian Mak, Matthew Ho, Jun Ye and Konstantina Koteva for fruitful discussions. The computations were performed on a Bull Qaptiva platform. This
work is part of HQI initiative (www.hqi.fr). This work is supported by France 2030 under the French National Research Agency grant numbers ANR-22-QMET-0002 (BACQ project) as part of the MetriQs-France program and grant number ANR-22-EXES-0013, as well as under the OECQ project financed by BPI France with France 2030 and Next Generation EU via France Relance), and by the National Research Foundation, Singapore through the National Quantum Office, hosted in A*STAR, under its Centre for Quantum Technologies Funding Initiative (S24Q2d0009). We acknowledge the support of the Singapore National Research Foundation (NRF) and French National Research Agency (ANR) joint project “QuRes” (Grant No. ANR-21-CE47-0019; NRF2021-NRF-ANR005).
\end{acknowledgments}

\appendix

  \begin{figure}
    \centering
    \includegraphics[width= \linewidth]{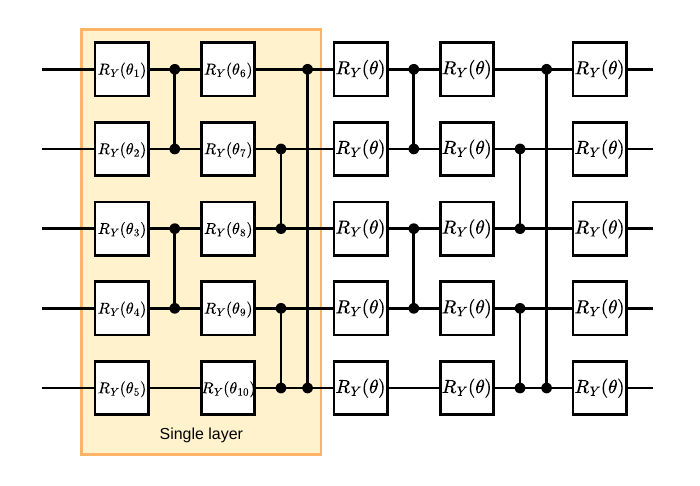}
    \caption{The abstract circuit of the RYA taken from \cite{Bravo-Prieto2020}. We do not transpile this circuit further, assuming all gates shown here: $R_Y (\theta), \, CZ$ are native. The number of gates is related to the number of layers as: $N_\textrm{g} = 25 N_\textrm{layers} + 5$, where each single qubit gate is counted once, and two qubit gate is counted twice, and identities are inserted for inactive qubits. Moreover, the final layer, before readout consists of $R_Y(\theta)$ gates on all qubits, contributing \textit{+5} term in the above relation.}
    \label{fig:RYA_abstract}
\end{figure}
\section{Validity of the scaling relations for other ansatzes}
\label{app:RYA}


\begin{figure}
    \centering
\includegraphics[width=\linewidth, trim=0cm 0cm 0cm 0cm, clip]{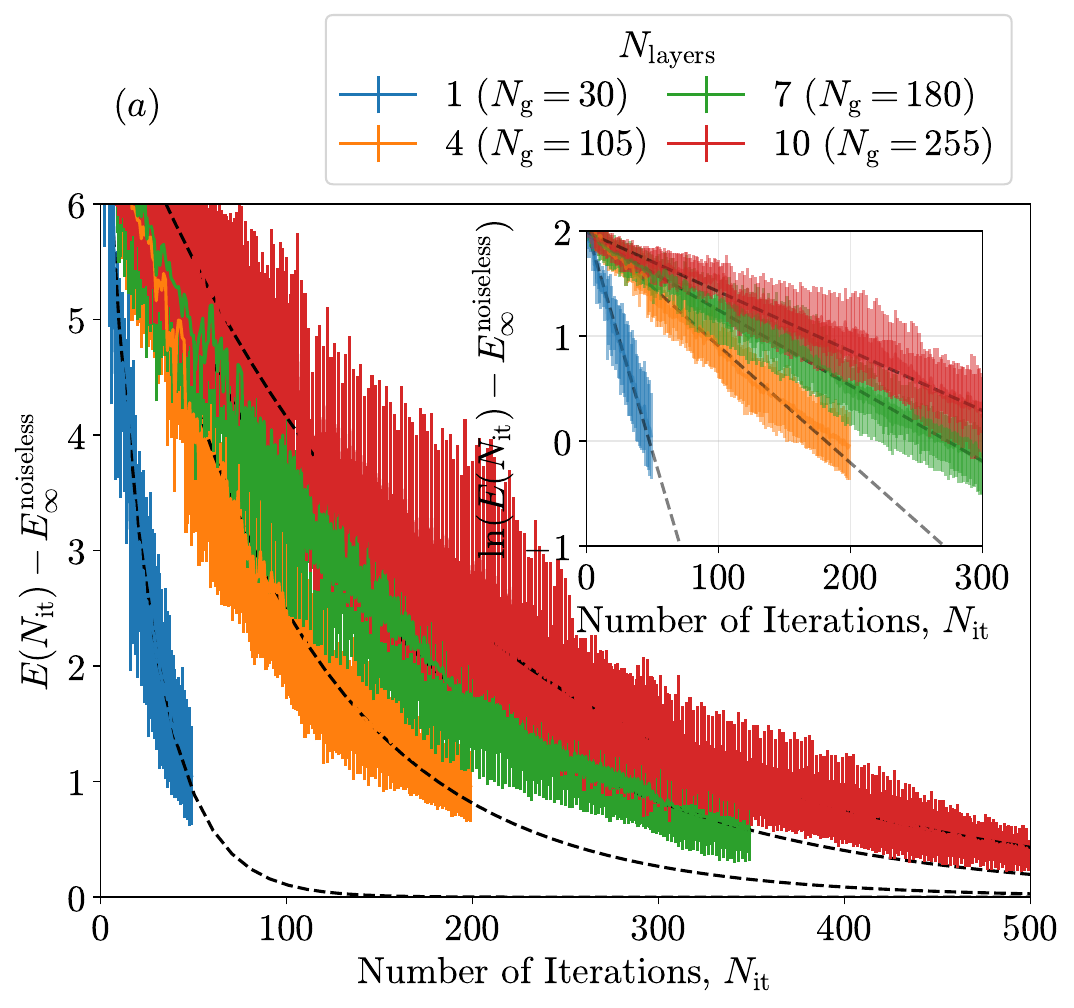}
\includegraphics[width=\linewidth, trim=0cm 0cm 0cm 0cm, clip]{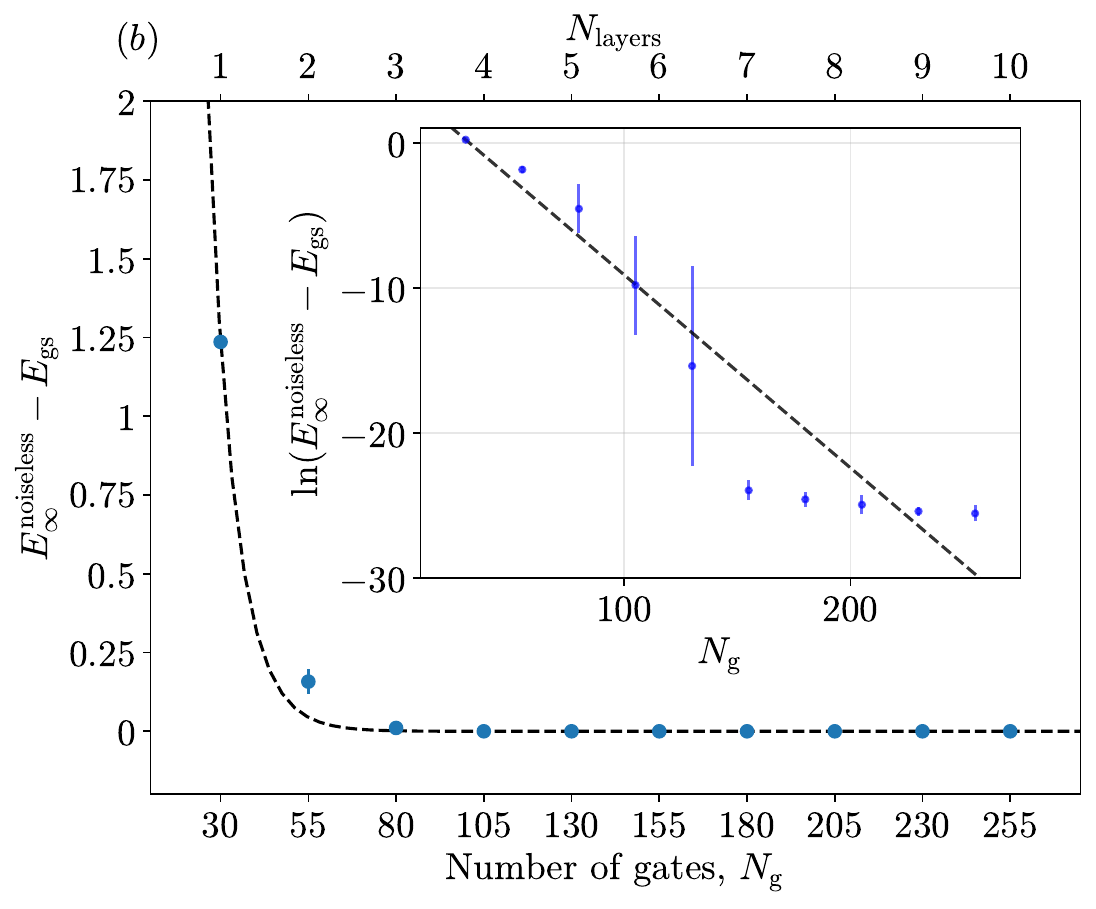}
\caption{The phenomenological scaling relations for RYA, used for extracting the fit parameters. The figures here are analogous to those shown for HVA in Fig.~\ref{fig:phen}, particularly Eq.~\eqref{eq:noiseless_soln} in (a), and Eq.~\eqref{eq:E_ng} in (b).}
    \label{fig:phen_RYA}
\end{figure}

  \begin{figure}
    \centering
    \includegraphics[width= \linewidth]{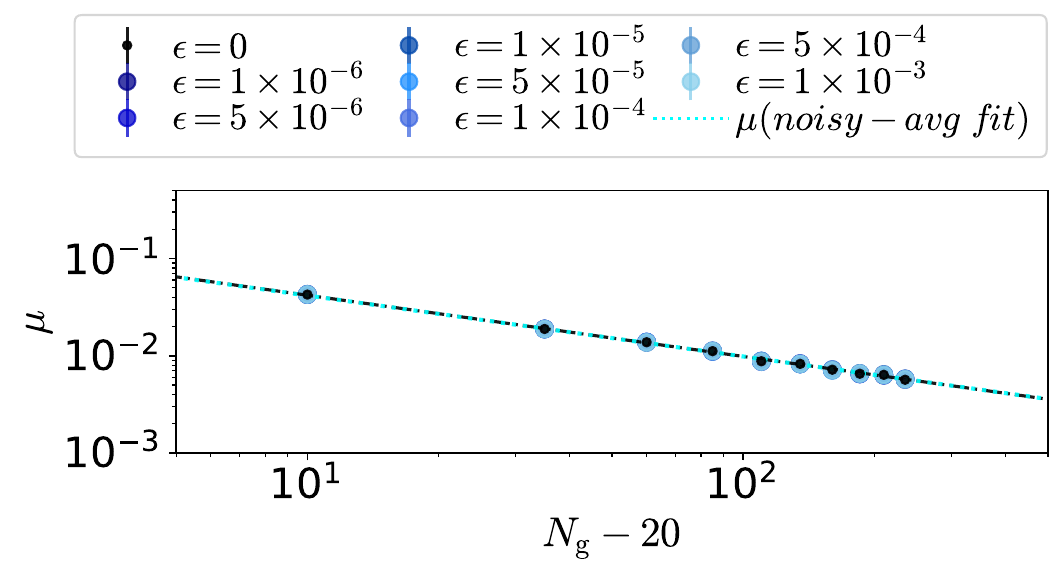}
    \caption{Phenomenological scaling relation for the convergence factor (with respect to iterations) $\mu$ as a function of the number of gates (offset by $N_\text{th} = 20$ here), as given in Eq.~\eqref{eq:gamma_fit}, shown here for RYA. }
    \label{fig:conv_rate_RYA}
\end{figure}

\begin{figure}
    \centering
    \includegraphics[width= \columnwidth]{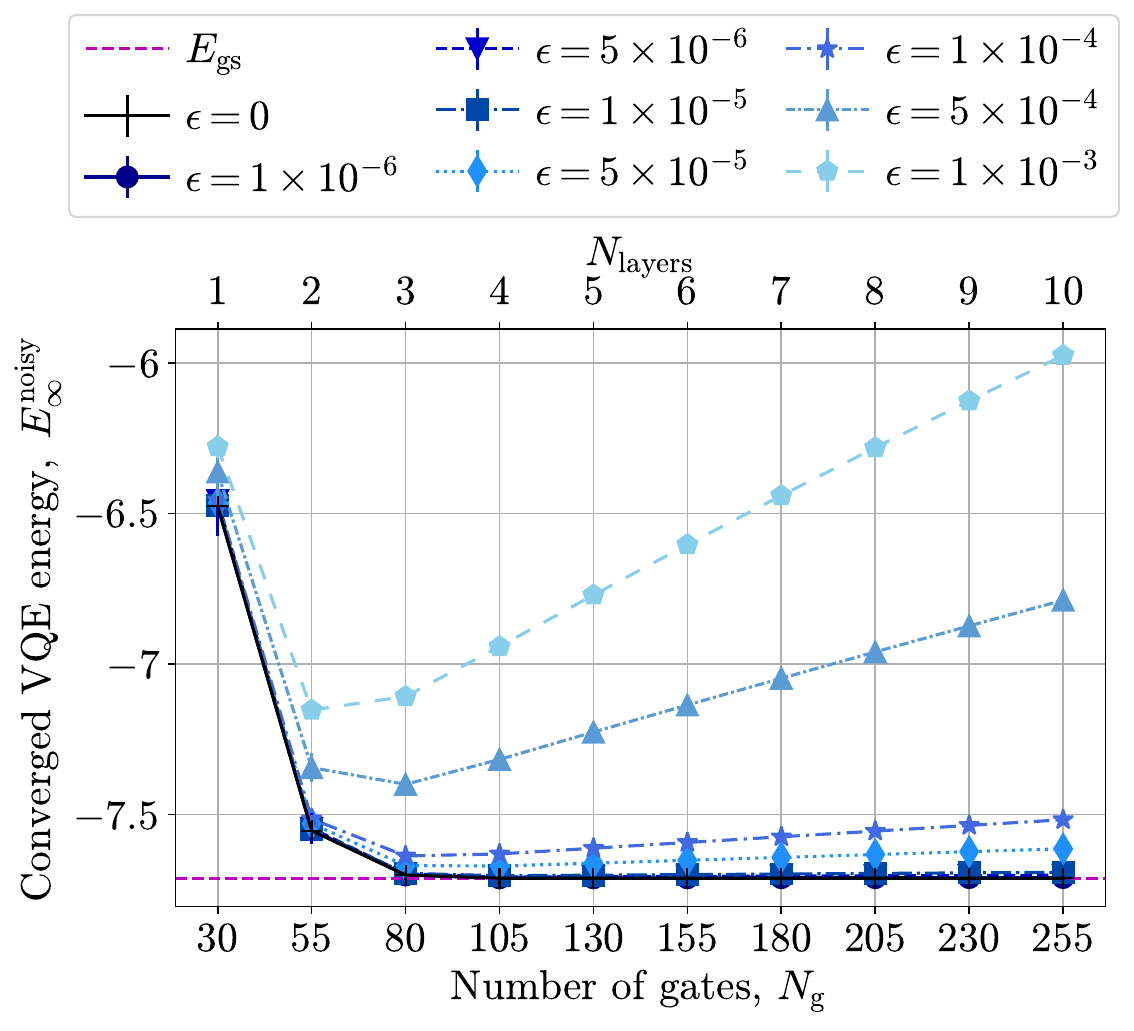}
\caption{The relationship between the number of gates and the converged energy for a noisy RYA based VQE algorithm. This figure is analogous to Fig.~\ref{fig:E_ng_tradeoff} which was based on HVA.}
    \label{fig:metric-resource_RYA}
\end{figure}
\setlength{\tabcolsep}{4pt} 
\renewcommand{\arraystretch}{1.5} 
\begin{table}
\begin{tabular}{l  c } 
 \hline
 \hline
 Parameter & Value (arb. units)  \\ [0.5ex] 
 \hline
  Number of random seeds & 50 \\ 
 
 Fitting range of $N_\textrm{it}$ for  E($N_\textrm{it}$)  & 50$N_\textrm{layers}$  \\ 
 
 $\mu_0$ & $0.18 \pm 0.01$  \\
 
  $\lambda$  & $0.63 \pm 0.01$  \\
 
 $\beta$ & $66.84 \pm 2.31$  \\ 
 
 $\alpha$(fiducial)& 7 \\
 
  $\kappa $ & 0.13  \\ 
 
 $N_\textrm{g}^\textrm{th} $ & 20  \\ 
 \hline 
 \hline
\end{tabular}
 \caption{Values for the phenomenological parameters obtained using fits corresponding to Eqs.~\eqref{eq:noiseless_soln}, \eqref{eq:gamma_fit} and\eqref{eq:E_ng} for the RYA with COBYLA optimizer.}
 \label{table:fit_param_RYA}
 \end{table}

The phenomenological scaling relations developed in Sec.~\ref{sec:VQE_MNR} were extracted from VQE algorithm simulations based on Hamiltonian Variational Ansatz. A natural question is whether these functional forms --- exponential convergence with iterations (Eq.~\ref{eq:noiseless_soln}), exponential improvement with circuit depth (Eq.~\ref{eq:E_ng}), and decay of the convergence rate (Eq.~\ref{eq:gamma_fit}) --- are specific to HVA or reflect more general features of variational optimization. To address this, we repeat the analysis using the RY ansatz (RYA) \cite{Bravo-Prieto2020} (see Fig.~\ref{fig:RYA_abstract}), a hardware-efficient ansatz with a different structure, for the same 5-qubit Heisenberg Hamiltonian in Eq.~\eqref{eq:heisen}. The RYA consists of layers of single-qubit $R_Y(\theta)$ rotations followed by a fixed entangling pattern (e.g., linear or circular CNOT chains), with no direct correspondence to the target Hamiltonian. In this case, we consider the $R_Y$ and $CZ$ gates to be native, and the number of gates is simply: $N_\textrm{g} = 5 + 25 N_\textrm{layers}$. We find that the scaling forms persist with different numerical coefficients, which attests to their generality. To this effect, we present the plots relevant to the scaling relations
 in Figs.~\ref{fig:phen_RYA}, \ref{fig:conv_rate_RYA}, \ref{fig:metric-resource_RYA}. The fit parameters extracted from these figures are provided in Table~\ref{table:fit_param_RYA}.

Note that the fit parameters, once extracted, allow us to explore the tradeoff between the algorithmic resources $N_\text{g}$ and $N_\textrm{it}$ as shown for HVA in Fig.~\ref{fig:energetics}. Furthermore, constrained optimization of resources can be done, as shown in Sec.~\ref{sec:tradeoffs} for HVA.

\section{Derivation of optimal $N_\text{g}$}
\label{app:Ngopt}

In this appendix, we derive the optimal number of gates per iteration, $N_\textrm{g}^\textrm{opt}$, that maximizes the algorithm's accuracy  for fixed total algorithmic cost $\Delta = N_{\textrm{it}} N_\textrm{g}$. We do this by taking Eq.~\eqref{eq:E_fixed_metric} for the algorithm's result, $E_\Delta(N_\textrm{g})$ and finding the $N_\textrm{g}$ that minimizes this result. As the algorithm's result is its prediction for the ground state of the Heisenberg-chain (and its prediction is always larger than the true result), minimizing  $E_\Delta(N_\textrm{g})$ corresponds to maximizing the algorithm's accuracy.

Hence the optimal $N_\textrm{g}^\textrm{opt}$ is given by finding the minimum of $E_\Delta(N_\textrm{g})$ in the usual manner, by using the conditions
\begin{eqnarray}
\left.\frac{\partial}{\partial N_\textrm{g}} E_\Delta(N_\textrm{g})\right|_{ N_\textrm{g}=N_\textrm{g}^\textrm{opt}} = 0,
~
\left.\frac{\partial^2}{\partial N_\textrm{g}^2} E_\Delta(N_\textrm{g})\right|_{ N_\textrm{g}=N_\textrm{g}^\textrm{opt}}> 0.~\nonumber\\
\end{eqnarray}
Taking the derivative of $E_\Delta(N_\textrm{g})$ in Eq.~\eqref{eq:E_fixed_metric} with respect to $N_\textrm{g}$,
setting it to zero and dividing by the positive factor $(1-\epsilon)^{N_\textrm{g}^\textrm{opt}}$, the stationarity condition becomes
\begin{subequations}
\label{eq:allsubequations-exact_stationarity}
\begin{eqnarray}
&\ln(1-\epsilon)\Big[\alpha e^{-\mu_0\Delta\,f(N_\textrm{g}^\textrm{opt})} + \beta e^{-\kappa N_\textrm{g}^\textrm{opt}} + E_\textrm{gs}\Big] = \nonumber \\
&\alpha \mu_0\Delta\,e^{-\mu_0\Delta\,f(N_\textrm{g}^\textrm{opt})}
f'(N_\textrm{g}^\textrm{opt}) + \beta \kappa e^{-\kappa N_\textrm{g}^\textrm{opt}}.
\label{eq:exact_stationarity}
\end{eqnarray}
where $f(N_\textrm{g}^\textrm{opt})= \big[N_\textrm{g}^\textrm{opt}( N_\textrm{g}^\textrm{opt}-N_\textrm{th})^\lambda\big]^{-1}$ and we define $f'(N_\textrm{g}) = \partial f(N_\textrm{g})/\partial N_\textrm{g}$,  so
\begin{eqnarray}
f'(N_\textrm{g}^\textrm{opt})
&=& - \frac{(1+\lambda)N_\textrm{g}^\textrm{opt} - N_\textrm{th}}{(N_\textrm{g}^\textrm{opt})^2\, (N_\textrm{g}^\textrm{opt}-N_\textrm{th})^{1+\lambda}}.
\label{eq:df_exact}
\end{eqnarray}
\end{subequations}
Eqs.~(\ref{eq:allsubequations-exact_stationarity}) give us a transcendental equation whose solution is the optimal value of $N_\textrm{g}$.
This transcendental equation requires a numerical solution, which is shown for various values of $\epsilon$ in Fig.~\ref{fig:Ng_vs_Delta}.

\subsection{Low-noise limit}

All simulations in this work employ low noise rates $\epsilon \in [10^{-6}, 10^{-3}]$, so it is natural to ask how the transcendental equation for optimal $N_\textrm{g}$, given in Eqs.~(\ref{eq:allsubequations-exact_stationarity}), simplifies in this regime.

Setting $\epsilon = 0$ in Eq.~\eqref{eq:exact_stationarity}, the left-hand side vanishes since $\ln(1) = 0$, yielding the following condition:
\begin{eqnarray}
\alpha\mu_0\Delta\, e^{-\mu_0\Delta\,f(N_\textrm{g})}\, f'(N_\textrm{g}) + \beta\kappa\, e^{-\kappa N_\textrm{g}} = 0.
\label{eq:noise_indep_balance}
\end{eqnarray}
Which we expect to also be approximately true for low-noise (small $\epsilon$).
The function $f(N_\textrm{g}) = [N_\textrm{g}(N_\textrm{g} - N_\textrm{th})^{\lambda}]^{-1}$ and its derivative $f'$ involve the threshold $N_\textrm{th} = 110$. Eq.~\eqref{eq:N_g} shows that, for the circuit structure in Fig.~\ref{fig:HVA}, $N_\textrm{g} = 150\, N_\textrm{layers}+15$, so whenever the optimal point has $N_\textrm{layers} \gtrsim 3$ (i.e.\ $N_\textrm{g} \gtrsim 465$) we have $N_\textrm{g} \gg N_\textrm{th}$, so $f(N_\textrm{g}) \approx N_\textrm{g}^{-(1+\lambda)}$ and  $f'(N_\textrm{g}) \approx -(1+\lambda)\,N_\textrm{g}^{-(2+\lambda)}$.
Substituting into Eq.~\eqref{eq:noise_indep_balance} gives the low-noise condition in Eq.~\eqref{eq:low_noise_balance}.

\subsection{The limit of large algorithmic resources ($\Delta \to \infty$)}
\label{subsec:largedelta}

Suppose that we have no limit on the algorithmic resources, so $\Delta$ can be as large as we like ($\Delta\to\infty$). Then we simply want to
get the highest algorithm accuracy in the limit of very large $\Delta$.

In this limit, it is always good to have infinitely many iterations of the quantum circuit to ensure complete convergence ($N_\textrm{it} \to \infty$), but it is counter-productive to have too many gate-operations in the quantum circuit, because the circuit will become too noisy.
This is what we see taking $\Delta \to\infty$ in Eqs.~(\ref{eq:allsubequations-exact_stationarity}) so that that equation reduces to the following equation for $N_\textrm{g}^\textrm{opt}$;
\begin{eqnarray}
\ln(1-\epsilon)\Big[\beta e^{-\kappa N_\textrm{g}^\textrm{opt}} + E_\textrm{gs}\Big] = \beta \kappa e^{-\kappa N_\textrm{g}^\textrm{opt}}.
\label{eq:stationarity_largeDelta_exact}
\end{eqnarray}
This is not a transcendental equation, it is simply an equation for $e^{-\kappa N_\textrm{g}^\textrm{opt}}$ and is 
easily solved to give
\begin{eqnarray}
N_\textrm{g}^\textrm{opt}(\Delta\to \infty) 
&=& 
\frac{1}{\kappa} \ln\left[\frac{\beta\,(\kappa -\ln(1-\epsilon))}{E_\textrm{gs}\,\ln(1-\epsilon)}  \right],
\label{eq:stationarity_largeDelta_exact-2}
\end{eqnarray}
where we recall that both $\ln(1-\epsilon)$ and $E_\textrm{gs}$ are negative. As our model assumes small $\epsilon$, we can write this as
\begin{eqnarray}
N_\textrm{g}^\textrm{opt}(\Delta\to \infty) 
&=& 
\frac{1}{\kappa} \ln\left[\frac{\beta\,(\kappa +\epsilon)}{|E_\textrm{gs}|\,\epsilon}  \right],
\label{eq:stationarity_largeDelta_exact-3}
\end{eqnarray}

Of course, Eqs.~(\ref{eq:stationarity_largeDelta_exact}-\ref{eq:stationarity_largeDelta_exact-3}) are the same equations that one would get for $N_\textrm{g}^\textrm{opt}$, if one directly optimizes the converged noisy energy, $E_\infty^\textrm{noisy}$ given by Eq.~\eqref{eq:E_conv_noisy} with Eq.~\eqref{eq:E_ng}.

Note that Eq.~\eqref{eq:stationarity_largeDelta_exact-3} exhibits a critical value of $\epsilon$, for which the logarithm in Eq.~\eqref{eq:stationarity_largeDelta_exact-3}) becomes negative. When $\epsilon$ is larger than this critical value, $\epsilon_\textrm{crit}$, the noise is so strong that optimal quantum circuit is the smallest possible circuit (a single-layer quantum circuit). This occurs when 
\begin{eqnarray}
\epsilon \ \geq \ \epsilon_\textrm{crit} \ \equiv \  \frac{\beta \kappa}{|E_\textrm{gs}|-\beta},
\end{eqnarray}
which means that the smaller $\beta$ and $\kappa$ are, the smaller the noise must be for it to be worth having quantum circuits with more than one layer. 

\section{Supplementary information for running VQE algorithm on \textit{fake} QPU}
\label{sec:sup_qpu}
Here we explain the steps involved in running the abstract circuit in Fig.~\ref{fig:HVA_abstract} on a \textit{fake} IBM QPU backend. We deliberately avoid pre-transpiling to a structured hardware-efficient layout, so that the compiler is free to optimize without an imposed intermediate circuit. Then,
device and qubit selection proceed in two stages. First, among available QPUs we compute the average two-qubit gate error over all nearest-neighbor pairs and select the device with the lowest value; in our case this is IBM \textit{Nighthawk}. Second, on the chosen device we identify five qubits with the lowest single-qubit gate errors, matching the system size of our toy model (a 5-qubit open-chain Heisenberg Hamiltonian). Compilation from the abstract circuit (Fig.~\ref{fig:HVA_abstract}) is handled by Qiskit's preset pass manager, which performs circuit compression, native gate-set conversion, and layout conformation. Circuit compression reduces the gate count and hence the accumulated error, but it is not an error mitigation technique in the formal sense.

\subsection{Error mitigation pipeline}
\label{sec:mitigation}

To escape the noise-dominated regime, we apply a three-stage mitigation pipeline at every optimization step --- that is, for every evaluation of $E(N_\textrm{it}, N_\textrm{g})$. The pipeline is implemented using the Mitiq library~\cite{mitiq} on top of Qiskit and combines noise tailoring with zero-noise extrapolation, applied in sequence from innermost (circuit-level) to outermost (estimator-level).

\textit{Pauli twirling.} Before each two-qubit gate in the compiled circuit (CZ being the native two-qubit gate on the Fake Nighthawk), we insert a random Pauli pair $P_i \otimes P_j$ before the gate and its conjugate after it, chosen so that the sandwich is logically equivalent to the identity~\cite{wallman2016noise,knill2004fault}. Averaged over many twirl realizations, this converts arbitrary coherent two-qubit noise into a stochastic Pauli channel. Twirling is performed because the subsequent zero-noise extrapolation step assumes that noise is incoherent and depolarizing-like; the twirling reduces any coherences, to make this assumption better and improves the reliability of the extrapolation. The twirled circuit is then passed through a basis-translation pass to ensure all inserted gates remain in FakeNighthawk's native basis.

\paragraph{Zero-noise extrapolation by gate folding.} Each two-qubit gate $G$ is replaced by three gates of the type $G\, G^\dagger\, G$, which leaves the intended unitary unchanged but amplifies the gate error by a so-called ``fold factor'' $c=3$~\cite{giurgica2020digital,temme2017error,li2017efficient}. The full circuit is executed without this replacement (error not amplified, $c=1$) and with this replacement (error amplified by a factor of $c=3$), yielding expectation values $E(c=1)$ and $E(c=3)$.

\paragraph{Linear (Richardson) extrapolation.} The zero-noise estimate is obtained by linear Richardson extrapolation of the two noise-amplified values,
\begin{equation}
    E(c=0) \;=\; \frac{3\,E(c=1) - E(c=3)}{2}.
    \label{eq:richardson}
\end{equation}

\paragraph{Outlook.} Uncertainty propagates in the error mitigation pipeline through shot-noise amplification during zero-noise extrapolation (Eq.~\eqref{eq:richardson}), which can be suppressed by increasing the number of seeds, twirl realizations, and shot budget. More sophisticated strategies --- higher-order zero-noise extrapolation~\cite{giurgica2020digital,temme2017error}, probabilistic error cancellation~\cite{temme2017error,vandenberg2023probabilistic}, Clifford data regression~\cite{czarnik2021error}, and symmetry verification~\cite{koczor2021exponential} --- would further reduce the residual non-depolarizing component and convert the present qualitative agreement into a quantitative one. We leave these techniques for future work.

\section{Self-consistent hardware workflow}
\label{app:outlook}

When working with real quantum hardware, noiseless reference data is unavailable and typical problems of interest cannot be solved classically. Our optimization must therefore be done using noisy data alone. The central question is how to extract the parameters of our scaling relations, such as Eq.~\eqref{eq:gen_soln}, without requiring noiseless data. 

This is straightforward if we assume that there is an error probability per gate operation of $\epsilon$, and that the noise is simple enough that the probability that a circuit of $N_\textrm{g}$ gates has {\it no errors} is $(1-\epsilon)^{N_\textrm{g}}$. 
This can be tested empirically by running a circuit with a simple known output, with various values of $N_\textrm{g}$. This circuit could be that of a randomized benchmarking protocol  \cite{RB_ref} or similar; in other words, it could be a circuit that does a series of gate-operations, followed by the reverse of those gate operations, so the error-free final state would be the same as the initial state.   This would confirm that the errors behave as $(1-\epsilon)^{N_\textrm{g}}$, and give a value of $\epsilon$. 
Once this factor of  $(1-\epsilon)^{N_\textrm{g}}$ is determined, one can divide through noisy data by this factor, and effectively eliminate the 
role of the noise in the scaling relations. One can then determine the other phenomenological parameters in the scaling relations ($\alpha$, $\beta$, $\mu_0$, $\lambda$, $\kappa$, and 
$N_\textrm{g}^{\textrm{th}}$) as we did in Sec.~\ref{sec:VQE_scaling} for noiseless data. 

The next question is why would noise in realistic hardware have such a simple scaling as $(1-\epsilon)^{N_\textrm{g}}$, since it relies on assuming that the noise is well-approximated by a depolarizing noise channel that is uncorrelated in time and uncorrelated between qubits. One reason is that 
Randomized Compiling (RC) is often used to ensure that the noise is of this type, because many error-estimation and error-mitigation techniques rely on this. Randomized compiling is a protocol that converts complex, hardware-specific noise into such a depolarizing channel.
It does this by inserting random gates into quantum circuits \cite{wallman2016noise, crosstalk_mitig}. 
Once the noise has been rendered depolarizing, cycle benchmarking or interleaved randomized benchmarking yields a per-gate $\epsilon$ which is rooted in a more realistic device error model consisting of crosstalk.
Randomized compiling and benchmarking thus play complementary roles: Randomized compiling  enforces the noise model that the scaling relations assume, while benchmarking independently measures the single parameter ($\epsilon$) of that model.

\bibliography{report_bib}
\end{document}